\documentclass[sn-mathphys-num]{sn-jnl}

\usepackage{graphicx}%
\usepackage{multirow}%
\usepackage{amsmath,amssymb,amsfonts}%
\usepackage{amsthm}%
\usepackage{mathrsfs}%
\usepackage[title]{appendix}%
\usepackage{xcolor}%
\usepackage{textcomp}%
\usepackage{manyfoot}%
\usepackage{booktabs}%
\usepackage{algorithm}%
\usepackage{algorithmicx}%
\usepackage{algpseudocode}%
\usepackage{listings}%
\usepackage{tabularx}

\usepackage{wasysym}

\newtheorem{theorem}{Result}

\begin{document}

\title[Article Title]{BRcal: An R Package to Boldness-Recalibrate Probability Predictions}

\author*[1]{\fnm{Adeline P.} \sur{Guthrie}}\email{apguthrie@vt.edu}

\author[1]{\fnm{Christopher T.} \sur{Franck}}\email{chfranck@vt.edu}

\affil[1]{\orgdiv{Department of Statistics}, \orgname{Virginia Tech}, \orgaddress{\street{250 Drillfield Drive}, \city{Blacksburg}, \postcode{24061}, \state{Virginia}, \country{USA}}}


\abstract{When probability predictions are too cautious for decision making, boldness-recalibration enables responsible emboldening while maintaining the probability of calibration required by the user. We formulate boldness-recalibration as a nonlinear optimization of boldness with a nonlinear inequality constraint on calibration.  We further show that recalibration based on the maximized linear log odds likelihood also maximizes the posterior probability of calibration. We introduce \textbf{BRcal}, an R package implementing boldness-recalibration and supporting methodology as recently proposed.  The \textbf{BRcal} package provides direct control of the calibration-boldness tradeoff and visualizes how different calibration levels change individual predictions.   We present a new real world case study involving housing foreclosure predictions.  The \textbf{BRcal} package is available on the Comprehensive R Archive Network (CRAN) (\url{https://cran.r-project.org/web/packages/BRcal/index.html}) and on Github (\url{https://github.com/apguthrie/BRcal}).}




\maketitle

\textbf{Acknowledgements}  The authors are thankful to Leidos for providing funding for this work. The authors would like to thank Matthew Keefe and Andrew McCoy for their insights and roles obtaining case study data for this line of research. The authors are grateful for the helpful comments from the editor and two referees.

\section{Introduction}\label{intro_rj}

Boldness-recalibration is a forecaster agnostic approach to responsibly embolden overly cautious probability predictions \cite{GuthrieFranck2024}.  When events occur at the rates that were predicted, the probability predictions are \textit{well calibrated}.   Calibration is important so that models provide decision makers with predictions that correspond to the actual rate of events.  However, overly cautious predictions tend to satisfy the literal definition of calibration (e.g., predicting sample proportion $\bar{y}$ for all events) but provide little insight on the forthcoming outcome.  Predictions that are further from $\bar{y}$ are more actionable. We call predictions that exhibit greater spread (and thus are typically further from $\bar{y}$) more \textit{bold} and we call the action of spreading predictions out \textit{emboldening}. To achieve bolder predictions, a sacrifice on calibration is typically required. Boldness-recalibration leverages this tradeoff between calibration and boldness by maximizing boldness at a specified level of calibration, subject to classification accuracy.  Rather than rely on calibration assessment techniques such as the Brier Score, Expected Calibration Error (ECE), and Log Loss, we employ a Bayesian model selection approach to quantify the probability of calibration via the posterior model probability. The posterior model probability of calibration is easily interpretable in the context of an individual forecaster, which is an advantage over these common calibration metrics.  The key feature of boldness-recalibration that sets it apart from most other calibration techniques is that it provides responsible emboldening of predictions rather than just improving calibration.  

In this paper, we present a clear demonstration of how to implement boldness-recalibration in practice and additional mathematical details pertaining to properties of maximum likelihood estimators (MLEs) in \cite{GuthrieFranck2024}.   Additionally, we present \textbf{BRcal}: an R package implementing boldness-recalibration and supporting methodology including calibration assessment via Bayesian model selection, LLO-adjustment via MLEs, and visualizations. 

Many R packages exist that implement recalibration techniques, calibration metrics, and visualizations such as reliability diagrams.  The \textbf{CalibrationCurves} \citep{CalibrationCurves-package} and \textbf{pmcalibration} \citep{pmcalibration-package} packages both implement and visualize Cox linear logistic calibration \citep{Dalton2013}. An implementation of isotonic regression for calibration \citep{ZadroznyElkan2002} can be found in the \textbf{rfUtilities} package \citep{rfUtilities-package}.  The \textbf{platt} package \citep{platt-package} focuses on Platt Scaling for calibration \citep{Platt2000}.  The \textbf{betacal} \citep{betacal-package} package provides Beta calibration with options for both 2- and 3-parameter approaches \citep{Kulletal2017}.  A bootstrap-based recalibration approach for a variety of regression models, including visualizations and scoring, can be found in the \textbf{rms} package.  The \textbf{braggR} package implements the approach for improving calibration via aggregation proposed in \cite{Satopaa2022}. The \textbf{scoring} \citep{scoring-package}, \textbf{scoringutils} \citep{scoringutils-package}, \textbf{DescTools} \citep{DescTools-package}, and \textbf{Yardstick} \citep{yardstick-package} packages primarily focus on scoring rules and other metrics including the Brier Score \citep{Brier1950}, Expected Calibration Error (ECE), and Log Loss.  Packages like \textbf{calibration} \citep{calibration-package}, \textbf{Calibtools} \citep{calibtools-package}, \textbf{CalibratR} \citep{CalibratR-package}, and \textbf{probably} \citep{probably-package} provide a suite of calibration tools including histogram binning \cite{ZadroznyElkan2001}, Bayesian Binning into Quantiles \citep{NaeiniCooperHauskrecht2015} and several of the previously mention recalibration techniques, scoring metrics, and visualizations.  None of these include boldness-recalibration.

The rest of this paper is organized as follows.  Section \ref{sec:backgound_rj} presents a brief and self contained review of the methods presented in \cite{GuthrieFranck2024}.  Section \ref{sec:methods_rj} presents key mathematical and optimization-related details of maximum likelihood estimation and boldness-recalibration. Section \ref{sec:BRcal} introduces the \textbf{BRcal} package by describing its core functions, implementation, and output from a familiar data example.  In Section \ref{sec:BRappl}, we demonstrate the capabilities of \textbf{BRcal} by presenting a new example related to housing foreclosure predictions.  Section \ref{sec:concl_rj} provides concluding remarks. 

\section{Background} \label{sec:backgound_rj}

 Let $\mathbf{x}$ be a vector of probability predictions from a single source (i.e., a single forecaster, or a single machine learning model, etc.) and let $\mathbf{y}$ be a vector of binary event outcomes.  We adopt the following likelihood function from \cite{GuthrieFranck2024}
\begin{equation}\label{likelihood_rj}
    \pi(\mathbf{y} | \delta, \gamma) = \prod_{i=1}^n c(x_i;\delta, \gamma)^{y_i} \left[1-c(x_i;\delta, \gamma)\right]^{1-y_i},
\end{equation}
where $c(x_i;\delta, \gamma)$ is the Linear Log Odds (LLO) recalibration function defined as 
\begin{equation}\label{llo_rj}
    c(x_i;\delta, \gamma) = \frac{\delta x_i^\gamma}{\delta x_i^\gamma + (1-x_i)^\gamma},
\end{equation}
and $\delta > 0$ and $\gamma \in \mathbb{R}$ are recalibration parameters \citep{Gonzalez1999, Turner2014}. Equation (\ref{llo_rj}) maps probability forecasts to the log odds scale, then shifts them by log $\delta$, scales by $\gamma$, and remaps back to the probability scale. Notice the shift is by $log(\delta)$ rather than $\delta$ because the shift occurs on the log odds scale. When we apply (\ref{llo_rj}) to prediction $x_i$ with some $\delta$ and $\gamma$, we call this \textit{LLO-adjustment}.

To assess calibration, we take an approximate Bayesian model selection-based approach comparing a well calibrated model ($\delta$ = $\gamma$ = 1) denoted by $M_c$, to an uncalibrated model ($\delta > 0$ and $\gamma \in \mathbb{R}$) denoted by $M_u$. When $\delta$ = $\gamma$ = 1 is appropriate, no shifting nor scaling of $\mathbf{x}$ is needed, which corresponds to the identify mapping of probability predictions to observed frequencies.  This corresponds directly to the frequency-based definition of calibration. The posterior model probability of $M_c$ is defined as
\begin{align} \label{pcalib_rj}
    P_{\mathbf{x}}(M_c|\mathbf{y}) = \frac{1}{1 + BF_{\mathbf{x}} \frac{P(M_u)}{P(M_c)}}.
\end{align}
where $BF_{\mathbf{x}}$ is the Bayes Factor comparing $M_u$ to $M_c$ defined as 
\begin{align}
    BF_{\mathbf{x}} = \frac{P_{\mathbf{x}}(\mathbf{y}|M_{u})}{P_{\mathbf{x}}(\mathbf{y}|M_c)}.
\end{align}
We approximate the Bayes Factor using the Bayesian Information Criteria (BIC) described in \cite{KassRaftery1995} and \cite{Kass1995Reference} such that \begin{align} \label{BF21}
    BF_{\mathbf{x}} \approx exp\left\{ -\frac{1}{2}(BIC_u - BIC_c) \right\}.
\end{align} 
Figure 5 of \cite{GuthrieFranck2024} shows that the BIC approximation for BF performs sensible updates to the posterior odds from the prior odds, even in sample sizes as small as 30.  In \eqref{BF21}, the BIC under the well calibrated model $M_c$ is defined as 
 \begin{align} \label{BIC1}
BIC_c &= - 2 \times log(\pi(\delta = 1, \gamma =1|\mathbf{x},\mathbf{y})).
\end{align}
The penalty term for number of estimated parameters is omitted in (\ref{BIC1}) as both parameters are fixed at 1 under $M_c$. The BIC under the uncalibrated model $M_{u}$ is defined as 
\begin{align} \label{BIC2}
BIC_u &= 2\times log(n) - 2\times log(\pi(\hat\delta_{MLE}, \hat\gamma_{MLE}|\mathbf{x},\mathbf{y})),
\end{align}
where $\hat\delta_{MLE}$ and $\hat\gamma_{MLE}$ are the maximum likelihood estimates (MLEs) for $\delta$ and $\gamma$ in (\ref{likelihood_rj}). The quantity $P_{\mathbf{x}}(M_c|\mathbf{y})$ is the posterior probability $\mathbf{x}$ is calibrated given observed outcomes $\mathbf{y}$ and is used to quantify calibration. \textit{MLE recalibration}, which is LLO-adjustment of $\mathbf{x}$ using the MLEs, produces maximally calibrated predictions $\mathbf{x}_{MLE}$ where $x_{i, MLE} = c(x_i;\hat\delta_{MLE}, \hat\gamma_{MLE})$.  {Section \ref{subsubsec:MLEmaximize} formally states this result.}


Maximally calibrated predictions are not necessarily bold enough for decision making.  For example, the Hockey study in \citep{GuthrieFranck2024} showed that even after MLE recalibration, the expert forecaster's predictions could be more bold while maintaining a high probability of calibration via boldness-recalibration.  On the other hand, random noise predictions were reigned in close to the base rate of 0.53 (i.e., not bold) to be maximally calibrated.  In general, predicting the base rate $\bar{y}$ for all events satisfies the definition of calibration, but lacks boldness. We define \textit{boldness} as the spread in probability predictions and measured by 
\begin{align} \label{sb_rj}
    s_b &= sd(\mathbf{x}),
\end{align}

\textit{Boldness-recalibration} finds values of $\delta$ and $\gamma$ that maximize $s_b$ subject to a user specified level of calibration denoted by $t$.  The parameters that achieve this goal are denoted by $\hat\delta_t$ and $\hat\gamma_t$ and are defined as 
\begin{align} \label{opt_rj}
    (\hat \delta_t,\hat \gamma_t) &= argmax_{( \delta, \gamma)}(s_b:P_{\mathbf{x'}}(M_c|\mathbf{y}, \delta,\gamma)\geq t),
\end{align}
where $x_i' = c(x_i; \delta, \gamma)$. Then $\mathbf{x}_t$ are called the $(100\cdot t)\%$ boldness-recalibrated predictions where $x_{i,t} = c(x_i; \hat\delta_t, \hat\gamma_t)$.

\section{Methods} \label{sec:methods_rj}

\subsection{Maximum Likelihood Estimation of $\delta$ and $\gamma$} \label{subsec:mlestimation}

There is no analytical solution for $\hat\delta_{MLE}$ and $\hat\gamma_{MLE}$ as the optimization is mathematically intractable.  Thus, all methods in this paper and in \citep{GuthrieFranck2024} rely on numerical optimization routines to acquire solutions for the MLEs.  Following common convention, we will minimize the negative log likelihood rather than maximize (\ref{likelihood_rj}) directly.  Additionally, we consider a reparameterization of the negative log likelihood in terms of $\tau = log(\delta) \in \mathbb{R}$.  Thus, the MLEs for $\tau$ and $\gamma$ can be written as
\begin{equation} \label{eq:mle_minim}
    (\hat \tau_{MLE},\hat \gamma_{MLE}) = argmin_{( \tau, \gamma)} \left(-\sum_{i=1}^n y_i log\left(c(x_i; exp(\tau), \gamma)\right) + (1-y_i) log\left(1-c(x_i; exp(\tau), \gamma)\right)\right),
\end{equation}
where the MLE for $\delta$ follows directly with $\hat \delta_{MLE} = exp(\hat \tau_{MLE})$. This reparameterization allows us to optimize in a completely unbounded parameter space, rather than bounding $\delta$ below at 0, which allows greater flexibility in choice of algorithm, as some routines do not allow bound constraints.  Additionally, we have found in practice that optimizing in terms of $\tau$ instead of $\delta$ is more efficient, which is explored further in the next section and Appendix \ref{app:optimalgcomps}.

\subsubsection{Optimization Routines for MLEs} \label{subsubsec:MLEroutines}

In this work, we primarily consider two optimization routines for acquiring MLEs: Nelder-Mead \citep{nelder-mead} and Limited Memory Broyden–Fletcher–Goldfarb–Shanno (L-BFGS) algorithm \citep{NocedalWright1998}.  We focus this section on these two algorithms because we have found them to be the most reliable in terms of convergence for MLEs in this work.

When comparing the performance of these two algorithms when used in this work, there is a trade off between efficiency, precision, and stability.  In terms of efficiency, L-BFGS has the benefit of being gradient-based. We have found that L-BFGS is 1.06 times faster than Nelder-Mead, on average.  However, L-BFGS is 2.00 times faster than Nelder-Mead when closed-form solutions for the gradients of the objective function are provided (on Ubuntu 22.04, R version 4.4.2).  The gradients of the objective function being maximized in \eqref{eq:mle_minim} can be found in Appendix \ref{app:optimalgcomps}.

The L-BFGS algorithm also benefits from slightly increased precision.  Table \ref{tab:optim_algs_intext} presents the standard deviation in the achieved MLEs ($\hat \delta_{MLE}$ and $\hat \gamma_{MLE}$) for the hockey data presented in \cite{GuthrieFranck2024} and in Section \ref{subsec:brcal_data} across a grid of different starting parameter values for L-BFGS and Nelder-Mead. While Nelder-Mead suffers from a slight deficit in efficiency and precision, we have found it to be more reliable than L-BFGS in that it is less prone to straying from the area in the parameter space where the optimum is located.  When L-BFGS iterates away from the optimum, it sometimes produces very extreme parameter values, which can cause the algorithm to abruptly terminate.   A more complete comparison of Nelder-Mead, L-BFGS, and a few other readily implemented algorithms can be found in Appendix \ref{app:optimalgcomps}.

\begin{table}[h]
\begin{tabular}{lll}
\hline
\textbf{Algorithm} &  \textbf{sd} $\mathbf{\hat \delta_{MLE}}$ & \textbf{sd} $\mathbf{\hat \gamma_{MLE}}$  \\
\hline\hline
L-BFGS      & $3.24 \times 10^{-6}$  & $1.67 \times 10 ^{-5}$ \\
Nelder-Mead & 0.0002                 & 0.0010 \\
\hline
\end{tabular}
\caption{Table summarizing variability in achieved MLEs for $\delta$ and $\gamma$ under optimization routines L-BFGS and Nelder-Mead in \texttt{optim()} for the \texttt{hockey} dataset.}
\label{tab:optim_algs_intext}
\end{table}

With these tradeoffs in mind, we primarily recommend Nelder-Mead for applications of this work.  Nelder-Mead is the default algorithm used in all cases where the MLEs are found in the \textbf{BRcal package}.  In cases where speed is of concern, we recommend trying L-BFGS with the closed-form gradients supplied, and in cases where this approach fails, return to Nelder-Mead.

\subsubsection{Maximizing $P_{\mathbf{x}'}(M_c|\mathbf{y})$} \label{subsubsec:MLEmaximize}

In Section \ref{sec:backgound_rj} and \cite{GuthrieFranck2024}, we claim that by LLO-adjusting a set of probability predictions by the MLEs for $\delta$ and $\gamma$, their calibration will be maximized.  This result can be more formally written as follows.

\begin{theorem}
Let $\mathbf{x}$ be a set of probability predictions with corresponding outcomes $\mathbf{y}$. Let $\mathbf{x}'$ be $\mathbf{x}$ under LLO-adjustment by some $\delta$ and $\gamma$.  Then under the BIC approximation of $BF_\mathbf{x'}$, $$(\hat\delta_{MLE}, \hat\gamma_{MLE}) = argmax_{( \delta, \gamma)}P_{\mathbf{x'}}(M_c|\mathbf{y}),$$  
meaning MLE recalibration maximizes the posterior probability of calibration.
\end{theorem}

In other words, the MLEs, which by definition maximize the LLO likelihood, also maximize the posterior probability of calibration as considered here. See Appendix \ref{app:MLEmaximize} for proof of this result.

\subsection{Boldness-Recalibration}

The problem of solving \eqref{opt_rj} to get the boldness-recalibration parameter estimates is also mathematically intractable.  In this paper, we formulate finding $\hat\delta_t$ and $\hat\gamma_t$ as a nonlinear constrained optimization problem. 

Again following common optimization conventions, we will minimize some objective function and enforce some constraint function be less than or equal to zero.  Since the goal of boldness-recalibration is to maximize boldness, measured by $s_b$, we will minimize negative $s_b$.  Thus, our objective function is as follows

\begin{align} \label{objective_rj}
    f(\delta, \gamma) &= -sd(\mathbf{x}') \\
    &= -\sqrt{\frac{\sum_{i=1}^n (x_i' - \bar{x}')^2}{n-1}}
\end{align}
where $\bar{x}'$ is the mean of $\mathbf{x}'$. Since boldness-recalibration requires that $(P_{\mathbf{x'}}(M_c|\mathbf{y})\geq t)$ we have the following constraint function after rearranging to follow the convention mentioned above 
\begin{align} \label{constraint_rj}
    g(\delta, \gamma) &= -(P_{\mathbf{x'}}(M_c|\mathbf{y})-t) \leq 0,
\end{align}
where $P_{\mathbf{x'}}(M_c|\mathbf{y})$ is the probability $\mathbf{x}'$ is calibrated given $\mathbf{y}$.  For clarity in describing and plotting the constraint, we refer to $P_{\mathbf{x'}}(M_c|\mathbf{y})$ as the \textit{constraint surface}, as it has the same shape and optimum as \eqref{constraint_rj}. Notice both the objective function $f(\delta, \gamma)$ and constraint function $g(\delta, \gamma)$ are nonlinear with respect to parameters $\delta$ and $\gamma$.

\begin{figure}[h!]
\begin{center}
\includegraphics[width=3in]{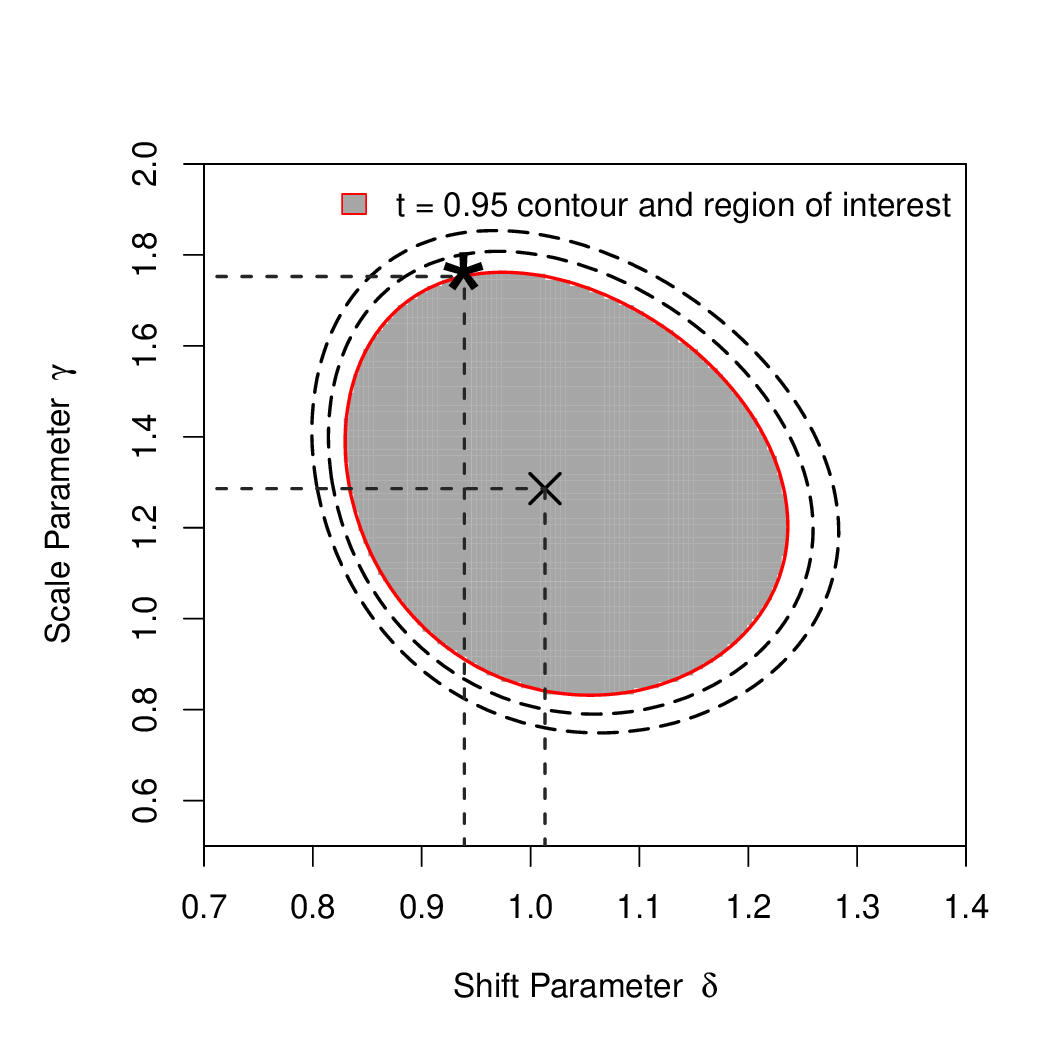}
\end{center}
  \caption{Schemas to visualize boldness-recalibration.  The left panel shows boldness in terms of spread in predictions, where each line corresponds to a prediction. The right panel shows a boldness-recalibration contour plot where the x, y, and z-axes correspond to shift parameter $\delta$, scale parameter $\gamma$, and $P_{\mathbf{x'}}(M_c|\mathbf{y})$, respectively. Contours correspond to $P(M_c|\mathbf{y})$ = 0.95 (solid red), 0.9 and 0.8 (dashed black). The shaded region corresponds to the constrained optimization region of interest under 95\% boldness-recalibration. The \textbf{$\APLstar$} on the 0.95 contour corresponds to ($\hat\delta_{0.95}$, $\hat\gamma_{0.95}$) such that the resulting probabilities have maximal spread subject to 95\% calibration. The $\times$ corresponds to ($\hat\delta_{MLE}$, $\hat\gamma_{MLE}$) such that the resulting probabilities under LLO-adjustment have maximal probability of calibration.}
  \label{fig:schema_ch3}
\end{figure}

This optimization approach is visualized in Figure \ref{fig:schema_ch3} which shows, as an example, the 95\% boldness-recalibration region of the parameter space.  The x and y-axes show values of $\delta$ and $\gamma$, respectively, and the z-axis is $P_{\mathbf{x'}}(M_c|\mathbf{y})$ (i.e., the constraint surface).  Rather than simply identify values for $\delta$ and $\gamma$ that maximize boldness metric $s_b$, we constrain our search to the region where the posterior model probability is at least $t=$0.95, which is represented by the shaded region in Figure \ref{fig:schema_ch3}. The \textbf{$\APLstar$} corresponds the values of $\delta$ and $\gamma$ within the shaded region that maximize spread.  Thus the values are denoted $\hat{\delta}_t$ and $\hat{\gamma}_t$ for $t=0.95$ in this example.  Plugging  $\hat{\delta}_t$, $\hat{\gamma}_t$, and the original predictions into \eqref{llo_rj} produces the 95\% boldness-recalibrated predictions.  The $\times$ corresponds to $\hat{\delta}_{MLE}$ and $\hat{\gamma}_{MLE}$.  

\subsubsection{Nonlinear Optimization Routines for Boldness-Recalibration} \label{subsubsec:nloptrroutines}

Since the constraint in \eqref{constraint_rj} is nonlinear with respect to $\delta$ and $\gamma$, we must consider different optimization routines than discussed in Section \ref{subsubsec:MLEroutines}.  In this work, we will consider routines that are readily available in \textbf{NLopt} \citep{nlopr-package}, a free open source nonlinear optimization library. 

  For all sets of predicted probabilities we encountered or generated for this work, the constraint surface (the posterior model probability) has had one unique maximum located at the MLEs (shown in Appendix \ref{app:MLEmaximize}).  Datasets include real-world case studies and simulated data intentionally generated with various levels of miscalibration.  The constraint region tends to take on a single connected region for fixed levels of $t$. Additionally, the objective function (the negative standard deviation) tends to have one unique minimum within the constraint region for fixed levels of $t$. Neither of these results are mathematically guaranteed, however, we have found them to be true in all cases we have explored, even in the most extreme cases of miscalibration.  A single connected region with a unique optimum is helpful for convergence, as it ensures that the optimizer will not get stuck in a disjoint, subpar region of the parameter space that satisfies the constraint.

\textbf{NLopt} includes both global and local optimization routines. Since we are typically  operating within a single constraint region with a single optimum, we need not consider global routines that explore the entire space. To ensure the optimizer can locate the region of interest, we recommend starting the optimization at the MLEs, since they are at the center of the region, as they maximize $P_{\mathbf{x'}}(M_c|\mathbf{y})$.

\textbf{NLopt} also includes both gradient-based and derivative-free routines. Given that the partial first derivatives of \eqref{objective_rj} and \eqref{constraint_rj} are available in closed-form, we will consider gradient-based optimization routines, as they are typically more efficient.  We define the Jacobians of \eqref{objective_rj} and \eqref{constraint_rj} in Appendices \ref{app:objgrad} and \ref{app:constrgrad}, respectively.  

  With this in mind, we chose to focus on local gradient based approaches of which there were three options that supported nonlinear inequality constraints: Sequential Least-Squares Quadratic Programming (SLSQP) \citep{slsqp1, slsqp2}, Augmented Lagrangian algorithm (AUGLAG) \citep{auglag1, auglag2}, and Method of Moving Asymptotes (MMA) \citep{MMA}.  Of these, we found AUGLAG produced the most stable results in terms of convergence in boldness-recalibration problems and is used throughout this work.   

The implementation of AUGLAG provided by \textbf{nloptr} reformulates the constrained optimization problem as a two-stage optimization.  The inner stage optimizes the objective function with a penalty for violating the constraint rather than use the constraint directly.  The outer stage governs the penalty.  If the inner stage returns a value that violates the constraint despite the penalty, the outer stage imposes a larger penalty. Note that since the inner stage is itself an unconstrained optimization routine, users may specify any optimization routine that supports nonlinear objective functions.  For the purpose of this package, we found that SLSQP as the inner optimization routine in AUGLAG provided the most stable results, thus SLSQP is set as the default inner optimization routine. For a more complete comparison of optimization routines and subroutines, see Appendix \ref{app:nloptralgcomps}.

The original proposal of boldness-recalibration in \cite{GuthrieFranck2024} does not explain how to practically implement the technique. Early implementations of boldness-recalibration used a grid search-based approach to find $\hat \delta_t$ and $\hat \gamma_t$.  However, this approach is inefficient and lacks precision. Compared to the original naive grid search approach, the nonlinear constrained optimization approach described above is 139 times faster, on average (on Ubuntu 22.04, R version 4.4.2).

\section{The BRcal package} \label{sec:BRcal}

The main contributions of the \textbf{BRcal} package are (i) boldness-recalibration, (ii) MLE recalibration, (iii) calibration assessment, and (iv) supporting visualizations.  The \verb|brcal()| function conducts boldness-recalibration at a user-specified level of calibration. The \verb|mle_recal()| function facilitates MLE recalibration.  The \verb|bayes_ms()| and \verb|llo_lrt()| functions performs Bayesian and frequentist approaches to calibration assessment, respectively.  The \verb|plot_params()| function provides a visualization for the constrained optimization via a color contour plot of the posterior model probability in (\ref{constraint_rj}).  The \verb|lineplot()| function provides a visualization for examining how individual predictions change under boldness-recalibration and MLE recalibration (optionally).  The package also provides a supporting function, \verb|LLO()|, for LLO-adjusting via any specified parameters.  

The \textbf{BRcal} package has three main dependencies on other R packages. The \verb|brcal()| function requires the \textbf{nloptr} package, which is the R implementation of the \textbf{NLopt} library, for nonlinear constrained optimization \citep{nlopr-package}. Additionally, the plotting functions make use of the \textbf{fields} \citep{fields-package} and \textbf{ggplot2} packages \citep{ggplot2-package}.  

The \textbf{BRcal} package is available on the Comprehensive R Archive Network (CRAN) (\url{https://cran.r-project.org/package=BRcal}) and on Github (\url{https://github.com/apguthrie/BRcal}).  For additional information and a helpful user guide with a tutorial, see the package vignette included on CRAN.

\subsection{Data} \label{subsec:brcal_data}

The \textbf{BRcal} package includes two built-in datasets called \verb|hockey| and \verb|foreclosure|.  

The \verb|hockey| dataset includes probability predictions of a home team win from FiveThirtyEight \citep{fivethirtyeight} and the winner of each game in the 2020-2021 National Hockey League (NHL) season. These data were used in \cite{GuthrieFranck2024}, and more information about the data can be found there.  All package demonstrations in this section utilize this data. The \verb|hockey| data frame includes four columns:
\begin{itemize}
    \item \verb|y| - integer; game result, 1 = home team win, 0 = home team loss
    \item \verb|x| - numeric; predicted probabilities of a home team win 
    \item \verb|rand| - numeric; randomly generated predicted probabilities of a home team win 
    \item \verb|winner| - factor; game result, ``home" = home team win, ``away" = home team loss.
\end{itemize}

The \verb|foreclosure| dataset includes probability predictions of foreclosure and the foreclosure status of 5,000 randomly selected housing transactions in 2010 from Wayne County, Michigan.  These data arose from a project that developed a monitoring strategy for the housing market using a spatially-adjusted model of foreclosure risk \citep{Keefe2017Monitoring}.  The probability of foreclosure was estimated using a mixture model of spatial kernel density estimates.  The model was trained during the first six months of 2005 using data from Wayne County, Michigan, which is a pre-Great Recession time period.  A case study analysis of this data is presented in Section \ref{sec:BRappl}. The \verb|foreclosure| dataframe includes three columns:
\begin{itemize}
    \item \verb|y| - integer; sale type, 1 = foreclosure, 0 = regular sale
    \item \verb|x| - numeric; predicted probabilities of foreclosure
    \item \verb|year| - numeric; year of observed foreclosure or regular sale
\end{itemize}


\subsection{The \texttt{brcal()} Function} \label{subsec:brcalf}

The core function of the \textbf{BRcal} package is the \verb|brcal()| function, which implements boldness-recalibration  using the nonlinear constrained optimization described in the previous section.  By default, \texttt{brcal()} performs 95\% boldness-recalibration (i.e., $t$=0.95).  Users can select a different minimum desired probability of calibration as seen in (\ref{opt_rj}) using argument \texttt{t}. Users should adjust this setting based on the level of risk they are willing to accept by sacrificing some calibration in favor of increased boldness.  The arguments for  \verb|brcal()| are summarized in Table \ref{tab:brcal_funct}.

\renewcommand{\arraystretch}{1.5}
\begin{table}[!h]
    \centering
\begin{tabular}{m{3.75 em} m{9.75 em}  m{26.5 em}  }
\hline
 \textbf{Function}  & \textbf{Arguments} & \textbf{Description}   \\
\hline\hline
 \verb|brcal()|  & \verb|x|  &   Vector of probability predictions  \\ 
        & \verb|y|  &   Vector of binary event outcomes  \\ 
        & \verb|t=0.95|  &   Level of user required calibration probability between 0 and 1\\ 
        & \verb|Pmc=0.5| & Prior model probability for calibrated model $M_c$\\
        & \verb|tau=FALSE| & Calculates in terms of $\tau = log(\delta)$ instead of $\delta$ when \verb|TRUE|\\
        & \verb|event=1| & Value in \verb|y| that represents an ``event''\\ 
        & \verb|start_at_MLEs=TRUE| &  Optimization starts at MLEs when \verb|TRUE| \\ 
        & \verb|x0=NULL|  &   Starting parameter values if \verb|start_at_MLEs = FALSE|\\
        & \verb|lb=c(1e-05, -Inf)|  &   Lower bounds for $\delta$ and $\gamma$\\
        & \verb|ub=c(Inf, Inf)|  &   Upper bounds for $\delta$ and $\gamma$\\
        & \verb|maxeval=500| & Max number of iterations for inner optimization\\
        & \verb|maxtime=NULL| & Max evaluation time (in seconds)\\
        & \verb|xtol_rel_outer=1.0e-6| & Tolerance for relative difference between iterations of outer optimization \\
        & \verb|xtol_rel_inner=1.0e-6| & Tolerance for relative difference between iterations of inner optimization\\
        & \verb|print_level=3|  &  Controls the amount of output from \verb|nloptr()| \\ 
        & \verb|epsilon=2.220446e-16|  &  Amount by which probabilities in \verb|x| are pushed away from 0 or 1 for numerical stability \\ 
        & \verb|opts=NULL|  &  List with options to be passed to \verb|nloptr()|   \\ 
        & \verb|optim_options=NULL|  &  List with options to be passed to \verb|optim()|   \\ 
 \hline
\end{tabular}
    \vspace*{0.3cm}
    \caption{Description of function arguments and their default values for the \texttt{brcal()} function.}
    \label{tab:brcal_funct}
\end{table}
\renewcommand{\arraystretch}{1.0}

Based on the discussion in Section \ref{subsubsec:nloptrroutines}, the \texttt{nloptr()} function uses the local gradient-based implementation of AUGLAG, with SLSQP as the inner routine, by default to find $\hat\delta_{t}$ and $\hat\gamma_{t}$ in \texttt{brcal()}. To leverage the closed-form derivative information, we supply the Jacobians of our objective and constraint functions from Appendices \ref{app:objgrad} and \ref{app:constrgrad} to \texttt{nloptr()} within \texttt{brcal()}. The \texttt{brcal()} function also uses \verb|optim()| to find $\hat\delta_{MLE}$ and $\hat\gamma_{MLE}$, however we primarily focus on how \verb|brcal()| uses the \verb|nloptr()| function in this section.  Discussion of \texttt{optim()} is reserved for Section \ref{subsec:calib_assess_fns}.

By default, \verb|brcal()| prints each step of the optimization process from \verb|nloptr()|. To suppress all output from \verb|nloptr()|, users can set \verb|print_level=0|.  Other options to minimize output can be found in the \verb|nloptr()| documentation and can be passed to \verb|brcal()| via \verb|print_level|.  Below we show an example where the full output from the first and last steps are shown, but the in-between stages are truncated by \verb|...| for readability.  

\begin{verbatim}
> br95 <- brcal(hockey$x, hockey$y)
iteration: 1
	x = (0.945397, 1.400573)
	f(x) = -0.123926
	g(x) = -0.048849
...
iteration: 146
	x = (0.872885, 1.958574)
	f(x) = -0.165339
	g(x) = -0.000000

> br95
nloptr

Call:
nloptr::nloptr(x0 = c(0.945397, 1.400573), eval_f = obj_f, eval_grad_f = obj_grad_f, 
    eval_g_ineq = constr_g, eval_jac_g_ineq = constr_grad_g, 
    opts = list(algorithm = "NLOPT_LD_AUGLAG", maxeval = 500, 
           maxtime = -1, xtol_rel = 1e-06, print_level = 3, 
           local_opts = list(algorithm = "NLOPT_LD_SLSQP", 
             eval_grad_f = obj_grad_f, eval_jac_g_ineq = constr_grad_g, 
             xtol_rel = 1e-06)), t = 0.95, tau = FALSE, Pmc = 0.5, 
    epsilon = 2.22044604925031e-16)


Minimization using NLopt version 2.7.1 

NLopt solver status: 4 ( NLOPT_XTOL_REACHED: Optimization stopped because 
xtol_rel or xtol_abs (above) was reached. )

Number of Iterations....: 146 
Termination conditions:  maxeval: 500	xtol_rel: 1e-06 
Number of inequality constraints:  1 
Number of equality constraints:    0 
Optimal value of objective function:  -0.165338762728956 
Optimal value of controls: 0.8728849 1.958574

$Pmc
[1] 0.5

$t
[1] 0.95

$BR_params
[1] 0.8728849 1.9585745

$sb
[1] 0.1653388

$probs
   [1] 0.5741797 0.7322471 0.5734560 0.5020176 0.5799966 0.3786775
   [7] 0.3228442 0.4505939 0.5180415 0.2241145 0.5229929 0.5265926
  ...
 [865] 0.8054830 0.7286748 0.2542999 0.4597869
\end{verbatim}

The \verb|brcal()| function returns a list with the following entries: (i) the information returned by the \verb|nloptr()| call in \verb|nloptr|, (ii) the prior model probability for $M_c$ in \verb|Pmc|, (iii) the user specified level of calibration desired in \verb|t|, (iv) the $(100\cdot t)\%$ boldness-recalibration parameters obtained from \verb|nloptr()| in \verb|BR_params|, (v) the maximum boldness achievable under $(100\cdot t)\%$ boldness-recalibration in \verb|sb|, and (vi) the resulting probabilities from recalibrating using the $(100\cdot t)\%$ boldness-recalibration parameters in \verb|probs|.

Notice the large amount of output from \verb|nloptr()| which is returned by \verb|brcal()| in \verb|nloptr|.  As with any optimization problem, it is important to verify convergence and adjust arguments as needed to achieve convergence. We include this copious output to allow users to check that the routine converged properly.  While this output shows the achieved boldness-recalibration estimates and standard deviation, we choose to return these separately in the list items \verb|BR_params| and \verb|sb|, respectively, for easier extraction.  The resulting set of boldness-recalibrated probabilities are also returned for easier extraction.

  The \verb|brcal()| function also provides, but does not require, full adjustment of the \verb|nloptr()| and \verb|optim()| routines.  The default arguments used in \verb|brcal()| have been effective in most applications explored by the authors.  A few options for stopping criteria, starting values, parameter bounds, and level of output are specified directly in \verb|brcal()|.  All other \verb|nloptr()| arguments, including the selection of optimization routine, can be optionally adjusted by passing them as a list to the \verb|opts| argument.  Similarly, all arguments in \verb|optim()| can be adjusted by passing them as a list to \verb|optim_options|.

With both an inner and outer optimization routine, stopping criteria are specified for each routine separately.  The \textbf{nloptr} package offers a variety of options including stopping based on the computation time, number of iterations, or change in function value from previous iteration.  By default, the \verb|brcal()| function stops the outer optimization when the relative difference in function value is $\leq 1.0e-6$ and stops the inner optimization when either the relative difference in function value is $\leq 1.0e-6$ or the number of evaluations surpasses 500.  Users of \verb|brcal()| are given the option to tweak any of these three options via parameters \verb|xtol_rel_outer|, \verb|xtol_rel_inner|, and \verb|maxevel|, or users can set the max evaluation time (in seconds) using \verb|maxtime|.  All other stopping criteria can be set using \verb|opts|.

For reasons discussed in Section \ref{subsubsec:nloptrroutines}, we set the default starting location of the \verb|nloptr()| optimizer to the MLEs for the original probability predictions.  We allow users to toggle this option via the \verb|start_at_MLEs| parameter.  By default, this parameter is set to \verb|TRUE|.  If set to \verb|FALSE|, users can specify an alternative starting value via argument \verb|x0|. Should users specify their own  \verb|x0|, the authors recommend imposing bounds on the parameter space using \verb|lb| and \verb|ub| to set lower and upper bounds, respectively, on $\delta$ and $\gamma$.

\subsection{Calibration Assessment} \label{subsec:calib_assess_fns}

While the goal of boldness-recalibration is to embolden when predictions are accurate, users can first assess calibration using the \verb|bayes_ms()| and \verb|llo_lrt()| functions.   The \verb|bayes_ms()| function uses the Bayesian model selection-based approach described in Section \ref{sec:backgound_rj}.  The \verb|llo_lrt()| function conducts the likelihood ratio test-based approach to assess calibration, details can be found in \cite{GuthrieFranck2024}.  Details of the arguments for \verb|bayes_ms()|, \verb|llo_lrt()|, and other supporting functions in the \textbf{BRcal} package can be found in Table \ref{tab:supp_functs}.

\renewcommand{\arraystretch}{1.5}
\begin{table}[!h]
    \centering
\begin{tabular}{m{4.75 em} m{9.75 em}  m{25.5 em}  }
\hline
 \textbf{Function}  & \textbf{Arguments} & \textbf{Description}   \\
\hline\hline
 \verb|bayes_ms()|    & \verb|x|  &   Vector of probability predictions  \\ 
 \verb|llo_lrt()|     & \verb|y|  &   Vector of binary event outcomes     \\
        & \verb|Pmc=0.5| & Prior model probability for calibrated model $M_c$ (\verb|bayes_ms()| only)\\
        & \verb|event=1| & Value in \verb|y| that represents an ``event'' \\
       & \verb|optim_details=TRUE| & Returns optimization details from \verb|optim()| when \verb|TRUE|\\
        & \verb|epsilon=2.220446e-16|  &  Amount by which probabilities in \verb|x| are pushed away from 0 or 1 for numerical stability \\ 
       & \verb|...| & Additional arguments to be passed to \verb|optim()|\\
 \hline
 \verb|mle_recal()|  & \verb|x|  &   Vector of probability predictions  \\ 
       & \verb|y|  &   Vector of binary event outcomes  \\ 
       & \verb|probs_only=TRUE| & Only returns the vector of recalibrated probabilities when TRUE \\
       & \verb|event=1| & Value in \verb|y| that represents and ``event'' \\
       & \verb|optim_details=TRUE| & Returns optimization details from \verb|optim()| when \verb|TRUE|\\
       & \verb|...| & Additional arguments to be passed to \verb|optim()|\\
 \hline
 \verb|LLO()|  & \verb|x| &  Vector of probability predictions \\
 & \verb|delta| & Value for $\delta$ to LLO-adjust \verb|x| by \\
 & \verb|gamma| & Value for $\gamma$ to LLO-adjust \verb|x| by\\
 \hline
\end{tabular}
    \vspace*{0.3cm}
    \caption{Description of the function arguments and default values for the \texttt{bayes\_ms()}, \texttt{llo\_lrt()}, \texttt{mle\_recal()}, and \texttt{LLO()} functions in the \textbf{BRcal} package.}
    \label{tab:supp_functs}
\end{table}
\renewcommand{\arraystretch}{1.0}

Both \texttt{bayes\_ms()} and \texttt{llo\_lrt()} provide numerical solutions for the MLEs for $\delta$ and $\gamma$ using the \texttt{optim()} function.  By default these functions use the \texttt{"Nelder-Mead"} routine as discussed in Section \ref{subsubsec:MLEroutines}.  While results are reported in terms of $\delta$, users can specify \texttt{tau=TRUE} to report results in terms of $\tau$ instead.  To make any adjustments to the call to \verb|optim()|, users can pass any arguments to \verb|optim()| using \verb|...|.

Notice from the output below that \verb|bayes_ms()| returns seven quantities in the form of a list.  The prior model probability for $M_c$ is returned in \verb|Pmc| to remind users what value was specified (0.5 by default).  List items \verb|BIC_Mc| and \verb|BIC_Mu| in (\ref{BIC1}) and (\ref{BIC2}) are the BICs for the calibrated model and uncalibrated model, respectively.  The Bayes factor comparing $M_u$ to $M_c$ in (\ref{BF21}) is returned in \verb|BF| and \verb|posterior_model_prob| is the posterior model probability of calibration from (\ref{pcalib_rj}). The MLEs for the vector \verb|x| are returned in \verb|MLEs|.  Additionally, the list returned by the call to \verb|optim()| to get the MLEs is returned in sublist \verb|optim_details|.  

\begin{verbatim}
> bt <- bayes_ms(hockey$x, hockey$y)
> bt
$Pmc
[1] 0.5

$BIC_Mc
[1] 1148.637

$BIC_Mu
[1] 1157.902

$BF
[1] 0.009730538

$posterior_model_prob
[1] 0.9903632

$MLEs
[1] 0.9453966 1.4005730

$optim_details
$optim_details$par
[1] 0.9453966 1.4005730

$optim_details$value
[1] 572.1846

$optim_details$counts
function gradient 
      63       NA 

$optim_details$convergence
[1] 0

$optim_details$message
NULL
\end{verbatim}

We now show the likelihood ratio test for calibration implemented in \verb|llo_lrt()|. This function returns the corresponding test statistic in \verb|test_stat|, p-value in \verb|pval|, MLEs in \verb|MLEs|, and the corresponding list from \verb|optim()| in \verb|optim_details|.  As expected, the MLEs and optimization details from \verb|llo_lrt()| match those from \verb|bayes_ms()|.

\begin{verbatim}
> llo_lrt(hockey$x, hockey$y)
$test_stat
[1] 4.267411

$pval
[1] 0.1183977

$MLEs
[1] 0.9453966 1.4005730

$optim_details
$optim_details$par
[1] 0.9453966 1.4005730

$optim_details$value
[1] 572.1846

$optim_details$counts
function gradient 
      63       NA 

$optim_details$convergence
[1] 0

$optim_details$message
NULL
\end{verbatim}

\subsection{MLE recalibration} \label{subsec:calib_assess}

The \textbf{BRcal} package also provides \verb|mle_recal()| to maximally calibrate predictions by LLO-adjusting via the MLEs.  This function returns $\mathbf{x}_{MLE}$, where $x_{MLE, i} = c(x_i; \hat{\delta}_{MLE}, \hat{\gamma}_{MLE})$, for a given set of probability predictions $\mathbf{x}$ and corresponding outcomes.  Users do not need to specify the MLEs as the function calculates this for them. 

\begin{verbatim}
> mle_recal(hockey$x, hockey$y)
$probs
  [1] 0.5633646 0.6814612 0.5628440 0.5117077 0.5675526 0.4223768
  [7] 0.3802169 0.4748424 0.5231574 0.3000678 0.5266953 0.5292678
  ...
[865] 0.7421488 0.6786381 0.3255808 0.4814569

$MLEs
[1] 0.9453966 1.4005730

$optim_details
$optim_details$par
[1] 0.9453966 1.4005730

$optim_details$value
[1] 572.1846

$optim_details$counts
function gradient 
      63       NA 

$optim_details$convergence
[1] 0

$optim_details$message
NULL

\end{verbatim}

By default, \verb|mle_recal()|  returns a list with (i) the vector of maximally calibrated predictions in \verb|probs|, (ii) the MLEs in \verb|MLEs| and (iii) the details from \verb|optim()| in \verb|optim_details|.  Should a user only want the vector of MLE recalibrated probabilities, they can set the option \verb|probs_only = TRUE|.  An example of when users may want this option is for assessing the calibration of the MLE recalibrated set.  By specifying \texttt{probs\_only = TRUE} users can pass the MLE recalibrated set directly from \texttt{mle\_recal()} to \texttt{bayes\_ms()} to get the maximum achievable probability of calibration for that set.  If the probabilities can only achieve a low probability of calibration, we would not recommend further emboldening the predictions, as they are likely poorly calibrated.

\subsection{LLO-Adjustment}
Should users want to LLO-adjust based on parameter values other than the MLEs or $(100\cdot t)\%$ boldness-recalibration parameters, we also provide the \verb|LLO()| function. The \verb|LLO()| function allows users to directly LLO-adjust a vector of probability predictions \verb|x| under any specified shift parameter \verb|delta| and scale parameter \verb|gamma|.  While most users of this package may not use the \verb|LLO()| function directly, there are some cases where users may want to do so. One example of this could be when you have MLEs for $\delta$ and $\gamma$ based on historical data and want to LLO-adjust unlabeled out-of-sample predictions. A user could use the LLO function to LLO-adjust the out-of-sample predictions by passing them to function argument \verb|x|, and passing the MLEs to arguments \verb|delta| and \verb|gamma|.  Another use case could be when you have boldness-recalibration parameter estimates from \verb|brcal()| but did not save the boldness-recalibrated probabilities.  Users could LLO-adjust their probabilities using the boldness-recalibration parameter estimates using \verb|LLO()| without needing to re-compute them using \verb|brcal()|, thus avoiding unnecessary computation time.  This is demonstrated below with the 95\% boldness-recalibration estimates saved in Section \ref{subsec:brcalf}.  Notice in the output below that we get the same boldness-recalibrated probabilities as we did using \verb|brcal()| in Section \ref{subsec:brcalf}. 

\begin{verbatim}
> LLO(hockey$x, delta=br95$BR_params[1], gamma=br95$BR_params[2])
  [1] 0.5741797 0.7322471 0.5734560 0.5020176 0.5799966 0.3786775
  [7] 0.3228442 0.4505939 0.5180415 0.2241145 0.5229929 0.5265926
  ...
[865] 0.8054830 0.7286748 0.2542999 0.4597869
\end{verbatim}

\subsection{Visualization Functions} \label{subsec:viz_imp}

The \textbf{BRcal} package provides two types of visualizations to conduct and assess the impact of boldness-recalibration: (i) a lineplot visualizing how individual predictions change under various LLO-adjustments and (ii) a color image and contour plot of the posterior model probability as a function of $\delta$ and $\gamma$.  Examples of both are provided in Figures \ref{fig:lineplot_rj} and \ref{fig:contourplot_rj} in Section \ref{sec:BRappl}.  This section describes the implementation details for these plot types. Description of the plotting function arguments and default values can be found in Table \ref{tab:plot_functs}.

\renewcommand{\arraystretch}{1.5}
\begin{table}[!h]
    \centering
\begin{tabular}{m{5.5 em} m{11.5 em}  m{23.25 em}  }
\hline
 \textbf{Function}  & \textbf{Arguments} & \textbf{Description}   \\
\hline\hline
 \verb|lineplot()|  & \verb|x|  &   Vector of probability predictions  \\ 
     & \verb|y|  &   Vector of binary event outcomes     \\
     & \verb|t_levels=NULL| &  Vector of desired level(s) of calibration at which to boldness-recalibrate\\
     & \verb|plot_original=TRUE| & Plots the original probabilities when \verb|TRUE|\\
     & \verb|plot_MLE=TRUE| & Plots the MLE Recalibrated probabilities when \verb|TRUE|\\
     & \verb|df=NULL| &  Dataframe of adjusted probabilities returned by previous call to \verb|lineplot()|\\
     & \verb|return_df=FALSE| &  Returns specially formatted dataframe of LLO-adjusted probabilities when \verb|TRUE| \\
     & \verb|thin_to=NULL| & Randomly thins to the specified number of observations \\
     & \verb|thin_prop=NULL| & Randomly thins to the specified proportion \\
     &\verb|thin_by=NULL| & Thins by taking every \verb|thin_by| observations \\ 
     &\verb|seed=0| & Seed used for random thinning for repeatability\\
     & \verb|ggpoint_options=NULL| & List of additional arguments passed to \verb|geom_point()|\\
     & \verb|ggline_options=NULL| & List of additional arguments passed to \verb|geom_line()|\\
 \hline
 \verb|plot_params()|  & \verb|x|  &   Vector of probability predictions  \\ 
     & \verb|y|  &   Vector of binary event outcomes     \\
     & \verb|z=NULL|  &  Matrix of posterior model probabilities across $k$ by $k$ grid of $\delta$ and $\gamma$ values returned by previous call to \verb|plot_params()| \\
     & \verb|t_levels=NULL| &   Vector of desired level(s) of calibration at which to plot contours\\
    &\verb|k=100| &  Grid of $\delta$ and $\gamma$ values is size $k$ by $k$\\
     & \verb|dlim=c(1e-04,5)| & Bounds for $\delta$ \\
     &\verb|glim=c(1e-04,5)| & Bounds for $\gamma$ \\
     &\verb|return_z=FALSE| & Returns matrix of posterior model probabilities when \verb|TRUE|\\
    & \verb|contours_only=FALSE| &  Only plots contours at levels in \verb|t_levels| when \verb|TRUE| with no grid cell coloring\\
    & \verb|imgplt_options=list(...)| & List of additional arguments passed to \verb|image.plot()| \\
    & \verb|contour_options=list(...)| & List of additional arguments passed to \verb|image.plot()|\\
 \hline
\end{tabular}
    \vspace*{0.3cm}
    \caption{Description of primary function arguments and their default values for the visualization functions (\texttt{lineplot()} and \texttt{plot\_params()}) in the \textbf{BRcal} package.  Note not all arguments are listed here.  See package documentation and vignette on CRAN for more details.}
    \label{tab:plot_functs}
\end{table}
\renewcommand{\arraystretch}{1.0}

\subsubsection{Lineplot} \label{subsec:lineplotfn}

\begin{figure}[h!]
\begin{center}
\includegraphics[width=3in]{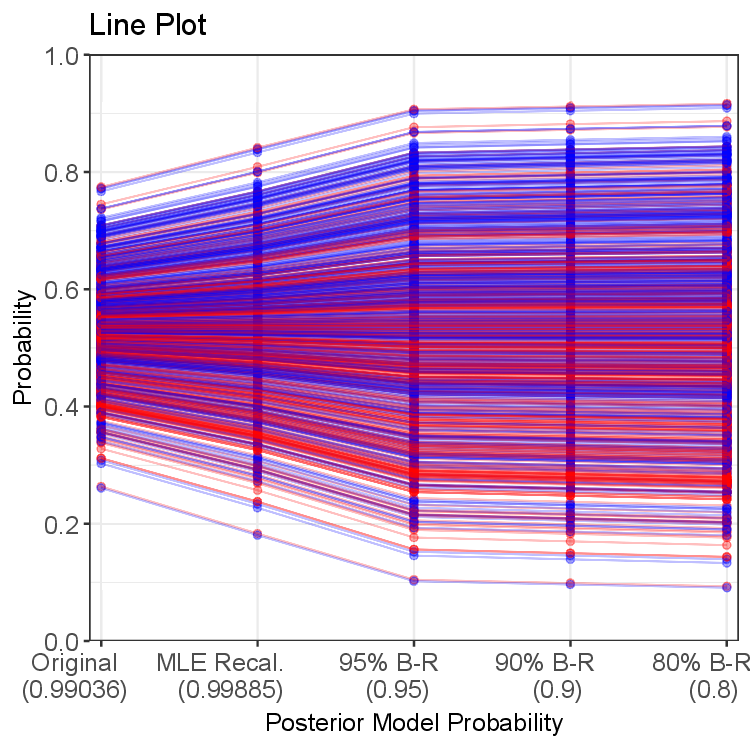}
\end{center}
  \caption{Demonstration of the \texttt{lineplot()} function on the \texttt{hockey} data.}
  \label{fig:hockeylineplot_ch3}
\end{figure}

The first visualization provided is a lineplot showing how probability predictions \textbf{x} change after LLO-adjustment implemented in the \verb|lineplot()| function and utilizes \textbf{ggplot2} graphics \citep{ggplot2-package}.  By default, only the original and MLE recalibrated probabilities are plotted.  To omit either the original or MLE recalibrated sets, users can set \verb|plot_original| or \verb|plot_MLE| to \verb|FALSE|.  To specify levels of boldness-recalibration, users can pass a vector of values for $t$ via \verb|t_levels|.   The resulting plot shows each specified set of probabilities side-by-side with lines connecting corresponding predictions. The code below generates the plot in Figure \ref{fig:hockeylineplot_ch3} as an example.  The lines are color coded based on the corresponding outcome, where blue represents $y_i$ = 1 (i.e., an ``event'') and red represents $y_i$ = 0 (i.e., a ``non-event'').  The goal is to visualize (i) the overall change in boldness  between the sets (e.g., spread increases/decreases under $95\%$ boldness-recalibration),  (ii) how individual predictions change (e.g., a prediction of 0.7 is adjusted to 0.8), and (iii) if predictions with high predicted probabilities typically result in events (e.g., point/lines for higher probabilities are typically blue, indicating an ``event''). 

\begin{verbatim}
> lineplot(hockey$x, hockey$y, t_levels=c(0.95, 0.9, 0.8))
\end{verbatim}

When the number of predictions for your lineplot is very large, plotting may be slow.  To ease the graphical burden, we offer three thinning options to reduce the amount of predictions that are visualized.  The \verb|thin_to| argument allows users to specify the number of predictions, $n_{thin}$, they would like to reduce their sample to.  In this case, \verb|lineplot()| will randomly select $n_{thin}$ predictions from \verb|x| and their corresponding outcomes in \verb|y|.  Using \verb|thin_prop| allows users to select the proportion of predictions to plot, which results in a random selection of that proportion size.  Lastly, specifying the \verb|thin_by| argument to some number $m$ where \verb|lineplot()| then selects every $m$ observation from the predictions.  Note that under all thinning strategies, all underlying calculations including the posterior model probabilities, boldness-recalibration parameters, and MLEs are conducted using the full set of probability predictions.  Thinning only reduces the number of observations seen on the plot.  

Once users have settled on the desired LLO-adjustments to be plotted (i.e., desired levels of boldness-recalibration, MLE recalibration, etc.), you can specify \verb|return_df=TRUE| to return the underlying dataframe of adjusted probabilities to avoid recalculation of these values.  This dataframe can then be reused by passing it to argument \verb|df|, indicating to \verb|lineplot()| that no recalculations are needed.  For example, it is useful to reuse \verb|df| so that users do not need to wait for the underlying boldness-recalibration calculations each time they make minor cosmetic adjustments to the plot, like modifying axes labels or font sizes.  

Users can make additional plotting adjustments to the underlying calls to \verb|geom_point()| and \verb|geom_line()| by passing the appropriate arguments as a list to \verb|ggpoint_options| and \verb|ggline_options()|, respectively.  Examples showing how to use these arguments can be found in the package vignette.

\subsubsection{Color Contour Plot} \label{subsubsec:colorcontourplot}

  The second visualization type is a color contour plot showing how the posterior model probability, $P_{\mathbf{x}'_{ij}}(M_c|\mathbf{y})$, changes with $\delta$ and $\gamma$. This visualization is implemented in the \verb|plot_params()| function and utilizes base R graphics and the \textbf{fields} package \citep{fields-package}. The resulting plot shows the posterior model probability surface over a $k$ by $k$ grid of $\delta$ and $\gamma$ values using a color scale to differentiate high and low values of $P_{\mathbf{x}'_{ij}}(M_c|\mathbf{y})$.  Additionally, users have the option to add contours at specified levels of calibration using \verb|t_levels|.  The goal of this is to visualize different values of $t$ at which a user may want to boldness-recalibrate, but \verb|plot_params()| does not perform boldness-recalibration nor identify $\hat{\delta}_t$ and $\hat{\gamma}_t$. We also recommend this plot be used to determine if the posterior model probability surface appears to be unimodal and take on one connected region, which can be important for optimization purposes.

\begin{figure}[h!]
\begin{center}
\includegraphics[width=3in]{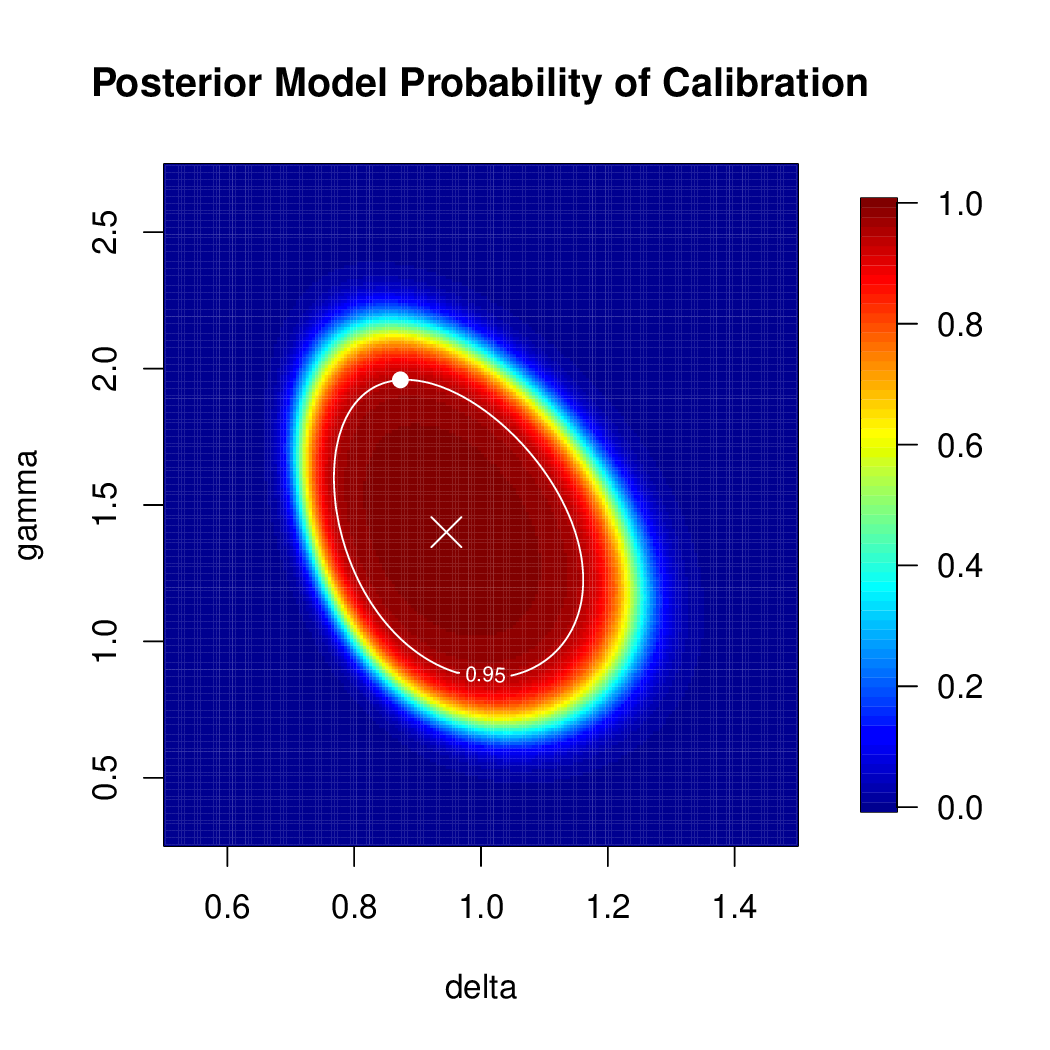}
\end{center}
  \caption{Demonstration of the \texttt{plot\_params()} function on the \texttt{hockey} data.}
  \label{fig:hockeycontours_ch3}
\end{figure}

  Arguments \verb|dlim| and \verb|glim| take the bounds for $\delta$ and $\gamma$ over which the user wishes to plot.  Argument \verb|k| is used to specify the density of the k by k grid of values.  In the code below, we examine \verb|k|=200 values of $\delta \in [0.5, 1.5]$ and 200 values of $\gamma \in [0.25, 2.75]$.  Since we are interested in 95\%, 90\%, and 80\% boldness-recalibration, we will also use \verb|t_levels=c(0.95, 0.9, 0.8)| to specify that we want contours drawn at those three levels. Additionally, since this plot uses base R graphics, we can add points denoting the boldness-recalibration parameters and MLEs using \verb|points()|.  Recall, these parameter estimates are saved in \verb|br95|, \verb|br90|, \verb|br80|, and \verb|bt|. The code below generates the example plot in Figure \ref{fig:hockeycontours_ch3}.

\begin{verbatim}

> plot_params(hockey$x, hockey$y, dlim = c(0.5, 1.5), glim = c(0.25, 2.75), 
+            k=200, t_levels = c(0.95))
> points(br95$BR_params[1], br95$BR_params[2], pch=19, col="white") 
> points(bt$MLEs[1], bt$MLEs[2], pch=4, col="white") 
\end{verbatim}

This plot can aid users in validating the MLEs and boldness-recalibration estimates for $\delta$ and $\gamma$.  For example, users should be concerned if the MLEs do no fall roughly in the middle of the dark red (high probability of calibration) region. Similarly, if the optimizer provides boldness-recalibration estimates that lie outside of the corresponding contour on the plot, users may need to make changes to the optimization routine to achieve convergence. While the optimization is not restricted to searching only directly on the contour of $P_{\mathbf{x'}}(M_c|\mathbf{y}) = t$, the optimal boldness-recalibration estimates typically fall on the upper edge of their corresponding contours. A reason for this is that $\gamma$ is the scale parameter (on the log odds scale).  Thus larger values of $\gamma$ tend to produce more bold (i.e., more spread out) predictions.  Additionally, this is a direct result of the tension between calibration and boldness. As we increase $\gamma$ beyond $\hat{\gamma}_{MLE}$, i.e., more away from maximal calibration marked by the $\times$ (the location of the MLEs), we gain boldness. Naturally, this places the solution to our optimization at the boundary of the constraint which is visualized by the contours.  Further intuition behind this tendency is provided in Appendix \ref{app:BRintuition}.

  Some care is required for the selection of the range of $\delta$ and $\gamma$ via arguments \verb|dlim| and \verb|glim|, and the size of $k$ via argument \verb|k|, which together determine the fineness of your grid of parameter values.  Small values of \verb|k| or misspecified parameter bounds may cause difficulties in detecting the region of non-zero $P_{\mathbf{x}'_{ij}}(M_c|\mathbf{y})$.  Large values of \verb|k| may cause a computational bottleneck as computation time increases with \verb|k|.  Appendix \ref{app:plt_params} presents recommendations for specifying \verb|k|, \verb|dlim|, and \verb|glim|.  Once users have settled on values of these arguments, you can specify \verb|return_z=TRUE| to return the underlying matrix of posterior model probabilities to avoid recalculation of these values.  This matrix can then be reused by passing it to argument \verb|z|, indicating to \verb|plot_params()| that no recalculations are needed.  Essentially, \texttt{plot\_params()} is using a streamlined implementation of the grid-search based approach for boldness-recalibration.  The improvement made to boost speed is discussed in Appendix \ref{app:pointslope}.

  Users can make additional plotting adjustments to the underlying calls to \verb|image.plot()| and \verb|countour()| by passing the appropriate arguments as a list to \verb|imgplt_options| and \verb|contour_options|, respectively.

\section{Foreclosure Monitoring Case Study} \label{sec:BRappl}

In this section we demonstrate \textbf{BRcal} using a case study related to the foreclosure risk modeling presented by Keefe et al. \citep{Keefe2017Monitoring}.  This example uses the \verb|foreclosure| dataframe built into the \textbf{BRcal} package, as described in Section \ref{subsec:brcal_data}.  We begin by loading the \textbf{BRcal} package and \verb|foreclosure| data and printing the first few rows of the data via the code below. 

\begin{verbatim}
> library(BRcal)
> data(foreclosure)
> head(foreclosure)
  y         x year
1 1 0.1155925 2010
2 1 0.3856198 2010
3 0 0.1273048 2010
4 1 0.4212137 2010
5 1 0.2623377 2010
6 1 0.5727253 2010
\end{verbatim}

With the data loaded, we can assess the calibration of the foreclosure predictions using the approximate Bayesian model selection approach via the \verb|bayes_ms()| function. We start with assessing calibration rather than boldness-recalibration, because it is good practice to understand the level of calibration we are starting at before making any adjustments. 

First, notice in the results below that the information returned in \verb|optim_details| indicates successful convergence on MLE values.  The MLE $\hat \delta_{MLE} = 11.109$ indicates that the predictions need substantial shifting (on the log odds scale) to maximize their calibration.  The MLE $\hat \gamma_{MLE} = 1.271$ indicates these predictions would benefit from a slight increase in boldness to maximize calibration.

\begin{verbatim}
> bt <- bayes_ms(foreclosure$x, foreclosure$y)
> bt
$Pmc
[1] 0.5

$BIC_Mc
[1] 8344.346

$BIC_Mu
[1] 4159.783

$BF
[1] Inf

$posterior_model_prob
[1] 0

$MLEs
[1] 11.109412  1.270519

$optim_details
$optim_details$par
[1] 11.109412  1.270519

$optim_details$value
[1] 2071.375

$optim_details$counts
function gradient 
      85       NA 

$optim_details$convergence
[1] 0

$optim_details$message
NULL
\end{verbatim}

While it is important to check the convergence of the MLEs to ensure reliable results, the primary goal of \verb|bayes_ms()| is to assess calibration.  The results above indicate that the foreclosure predictions are not well calibrated, as their posterior probability of calibration is 0.000.

We now show the likelihood ratio test for calibration implemented in \verb|llo_lrt()|.  As expected, the MLEs and optimization details from \verb|llo_lrt()| match those from \verb|bayes_ms()|.  Additionally, the p-value is near zero, also indicating these predictions are not well calibrated.

\begin{verbatim}
> llo_lrt(foreclosure$x, foreclosure$y)
$test_stat
[1] 4201.597

$pval
[1] 0

$MLEs
[1] 11.109412  1.270519

$optim_details
$optim_details$par
[1] 11.109412  1.270519

$optim_details$value
[1] 2071.375

$optim_details$counts
function gradient 
      85       NA 

$optim_details$convergence
[1] 0

$optim_details$message
NULL
\end{verbatim}

To quantify the maximum achievable probability of calibration of these predictions, we can first MLE recalibrate using \texttt{mle\_recal()} and then assess the calibration of the MLE recalibrated set using  \texttt{bayes\_ms()}. This is executed in the code below and the output indicates that the maximum achievable probability of calibration of these predictions is 0.9998.

\begin{verbatim}
> mle_probs <- mle_recal(foreclosure$x, foreclosure$y, probs_only = TRUE)
> bayes_ms(mle_probs, foreclosure$y)$posterior_model_prob
[1] 0.9998
\end{verbatim}

Now that we have a baseline value of calibration, we can explore 95\%, 90\%, and 80\% boldness-recalibration using the \verb|brcal()| function.  Only the truncated output from 95\% boldness-recalibration is shown below.

\begin{verbatim}
> br95 <- brcal(foreclosure$x, foreclosure$y)
iteration: 1
	x = (11.109412, 1.270519)
	f(x) = -0.179135
	g(x) = -0.049800
...
iteration: 201
	x = (12.134803, 1.411657)
	f(x) = -0.199055
	g(x) = -0.000000

> br90 <- brcal(foreclosure$x, foreclosure$y, t=0.9, print_level=0)
> br80 <- brcal(foreclosure$x, foreclosure$y, t=0.8, print_level=0)

> br95
$nloptr

Call:

nloptr::nloptr(x0 = c(11.109412, 1.270519), eval_f = obj_f, eval_grad_f = 
    obj_grad_f, eval_g_ineq = constr_g, eval_jac_g_ineq = constr_grad_g, 
    opts = list(algorithm = "NLOPT_LD_AUGLAG", maxeval = 500, 
        maxtime = -1, xtol_rel = 1e-06, print_level = 3, local_opts = 
        list(algorithm = "NLOPT_LD_SLSQP", eval_grad_f = obj_grad_f, 
        eval_jac_g_ineq = constr_grad_g, xtol_rel = 1e-06)), 
    t = 0.95, tau = FALSE, Pmc = 0.5, epsilon = 2.22044604925031e-16)
    
Minimization using NLopt version 2.7.1 

NLopt solver status: 4 ( NLOPT_XTOL_REACHED: Optimization stopped 
because xtol_rel or xtol_abs (above) was reached. )

Number of Iterations....: 201 
Termination conditions:  maxeval: 500	xtol_rel: 1e-06 
Number of inequality constraints:  1 
Number of equality constraints:    0 
Optimal value of objective function:  -0.199055078798425 
Optimal value of controls: 12.1348 1.411657

$Pmc
[1] 0.5

$t
[1] 0.95

$BR_params
[1] 12.134803  1.411657

$sb
[1] 0.1990551

$probs
   [1] 0.4069886 0.8627809 0.4448838 0.8856910 0.7381995 0.9483221
   [7] 0.9313465 0.4707687 0.9271990 0.6411276 0.4741822 0.8375914
  ...
 [997] 0.9028353 0.8908336 0.3073575 0.9500874
 [ reached getOption("max.print") -- omitted 4000 entries ]

\end{verbatim}

Notice above that the optimization routine terminated because our stopping condition of \verb|xtol_rel = 1e-06| was met, meaning the algorithm stops when the change in parameter values from one iteration to another, relative to the size of the parameter values, is smaller than $1\times10^{-6}$.  We consider this successful convergence.

The estimated 95\% boldness-recalibration parameters are $\hat \delta_{0.95} = 12.135$ and $\hat \gamma_{0.95}=1.412$. Additionally, we see that the maximum achievable boldness under the constraint of $P_{\mathbf{x}'}(M_c|\mathbf{y}) \geq 0.95$ is $s_b=0.199$. Note that on average, 95\% boldness-recalibration on this data takes about 33 seconds to run (on Ubuntu 22.04, R version 4.4.2).

\begin{figure}[h!]
\begin{center}
\includegraphics[width=3in]{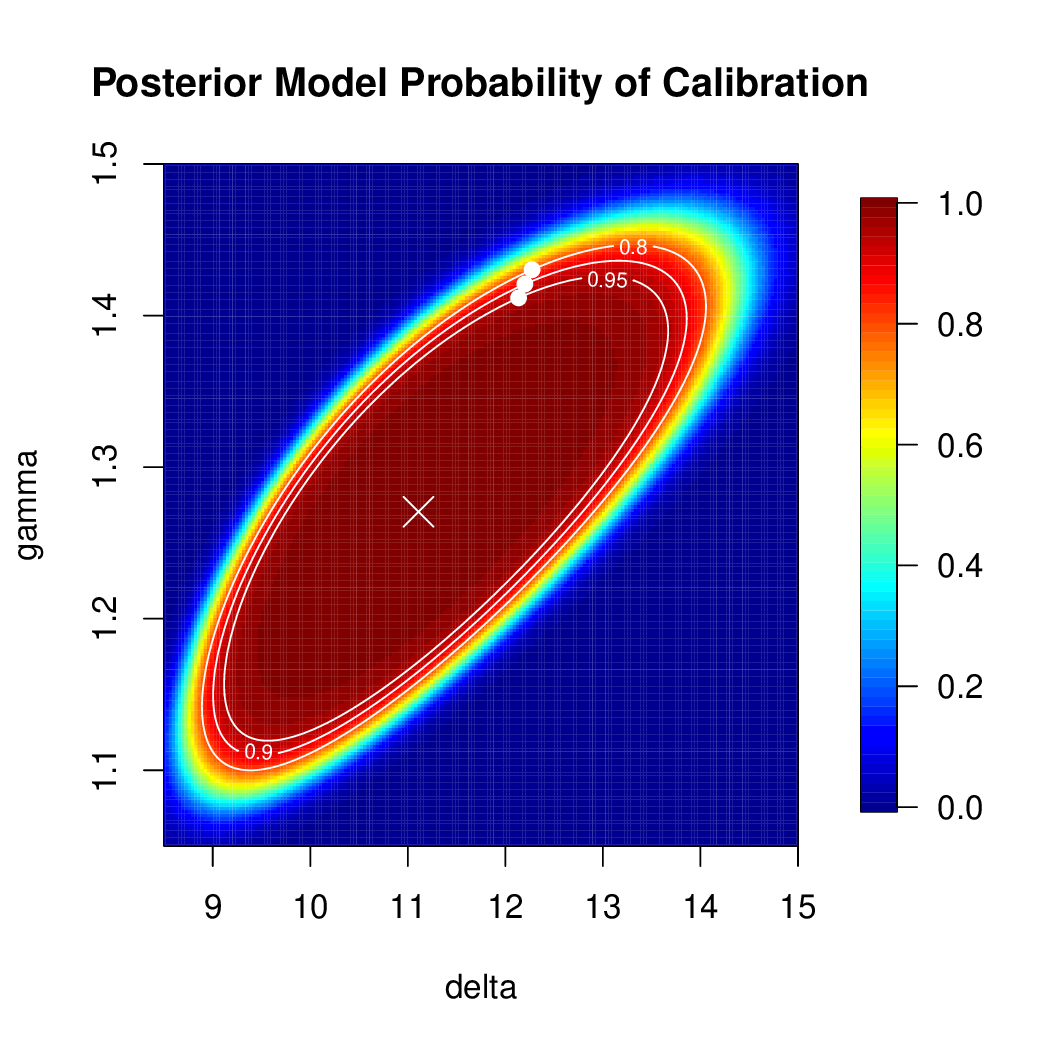}
\end{center} 
  \caption{Color contour plot demonstrating the \texttt{plot\_params()} function. The points in white along the contours denote the boldness-recalibration parameters.  The $\times$ denotes the MLEs. The color of each grid cell represents $P_{\mathbf{x'}}(M_c|\mathbf{y})$, where $\mathbf{x'}$ is LLO-adjusted by the corresponding $\delta$ and $\gamma$ on the $x$ and $y$-axes, respectively.}
\label{fig:contourplot_rj}
\end{figure}

 The estimated MLEs, 95\%, 90\%, and 80\% boldness-recalibration parameters and their achieved $P_{\mathbf{x}'}(M_c|\mathbf{y})$ are visualized in Figure \ref{fig:contourplot_rj}, generated by the code below.  This plot supports our conclusion that both the boldness-recalibration and MLE optimizations converged appropriately, as the MLEs fall directly at the optimum of $P_{\mathbf{x}'}(M_c|\mathbf{y})$ (where the graph is the deepest shade of red) and the boldness-recalibration estimates fall along the upper boundary of their respective contours.  
 
\begin{verbatim}
> plot_params(foreclosure$x, foreclosure$y,
+              dlim = c(8.5, 15), glim = c(1.05, 1.5), k=200,
+              t_levels = c(0.95, 0.9, 0.8))
> # add 95% B-R params
> points(br95$BR_params[1], br95$BR_params[2], pch=19, col="white") 
> # add 90% B-R params
> points(br90$BR_params[1], br90$BR_params[2], pch=19, col="white") 
> # add 80% B-R params
> points(br80$BR_params[1], br80$BR_params[2], pch=19, col="white")
> # add MLEs
> points(bt$MLEs[1], bt$MLEs[2], pch=4, col="white", cex=2)      
\end{verbatim}

\begin{figure}[h!]
\begin{center}
\includegraphics[width=3in]{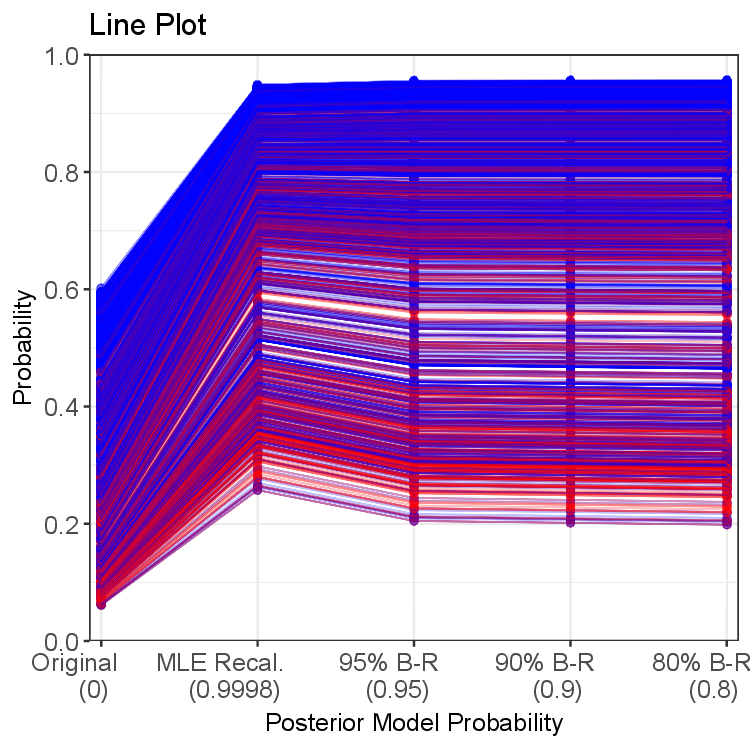}
\end{center}
  \caption{Demonstration of the \texttt{lineplot()} function visualizing how predictions change from varying recalibration strategies. The first column of points is the original set of probability predictions.  The second column are the MLE recalibrated predictions (i.e., after recalibrating with $\hat\delta_{MLE}$ and $\hat\gamma_{MLE}$).  The last three columns are the predictions after 95\%, 90\%, and 80\% boldness-recalibration (i.e., $\mathbf{x}_{0.95}$, $\mathbf{x}_{0.90}$, and $\mathbf{x}_{0.80}$) respectively. A line is used to connect each original prediction to where it ends up after each recalibration procedure. Blue points and lines correspond to predictions for which there was a foreclosure (i.e., ``events'').  Red corresponds to transactions in which there was no foreclosure (i.e., ``non-events''). Achieved $P_{\mathbf{x}'}(M_c|\mathbf{y})$ are reported on the x-axis in parentheses. }
  \label{fig:lineplot_rj}
\end{figure}

Below we visualize how MLE recalibration and boldness-recalibration (for t=0.95, 0.9, and 0.8) change the individual probabilities themselves using \verb|lineplot()|. The code below generates the plot shown in Figure \ref{fig:lineplot_rj}.  Given the size of this data, the plot may take a few minutes to generate.  In practice, users can leverage the strategies discussed in Section \ref{subsec:lineplotfn} for reducing computation time. 

\begin{verbatim}
> lineplot(foreclosure$x, foreclosure$y, t_levels=c(0.95, 0.9, 0.8))
\end{verbatim}

There are a few key takeaways from the visualization in Figure \ref{fig:lineplot_rj}.  First, the original foreclosure predictions are relatively accurate in terms of classification accuracy as more of the observations are colored blue (corresponding to foreclosure) for higher probabilities and more are colored red (corresponding to non-foreclosure) for lower probabilities.  In fact, these predictions achieve an area under the ROC curve (AUC) of 0.8.  Note that accuracy as measured by AUC does not change under any LLO-adjustment, since LLO-adjustment does not change the order of the predictions and AUC considers every possible threshold in the data.  However, measures of accuracy that require a single fixed threshold to determine which observations should be classified as an ``event'' or ``non-event'' may change under LLO-adjustment, as the adjustment may push predictions to the opposite side of the threshold.  Second, the original predictions are substantially biased, as they consistently underestimate the rate of foreclosure, which on average in this dataset set is 0.797.  This bias is most likely due to using a model trained on pre-recession data from 2005 to estimate foreclosure rates for transactions from post-recession in 2010. MLE recalibration corrects this bias by shifting all predictions towards the observed base rate of foreclosure of 0.797.  This is further reflected by the large value of $\hat \delta_{MLE}$, as larger values of $\delta$ indicate greater shifts in the predictions.  The mean prediction shifts from 0.38 to 0.80. Finally, these predictions can also benefit from being spread out under MLE recalibration, meaning we can both increase calibration and boldness at the same time, producing predictions more useful for decision making. The spread in the predictions increases from 0.16 in the original set (range: 0.06, 0.6), to 0.18 in the MLE recalibrated set (range: 0.26, 0.95).  This increase in boldness arises from the fact that $\hat \gamma_{MLE} >1$.  Then, with a slight relaxation of calibration probability from the maximum achievable, 0.9998, to 0.95, the spread in predictions increases further to 0.20. This is the maximum boldness achievable while maintaining a calibration probability of 0.95. The further increase in spread is also reflected in the fact that $\hat \gamma_{0.95} > \hat \gamma_{MLE} >1$.

\section{Conclusion} \label{sec:concl_rj}

In this paper, we formulate boldness-recalibration as a nonlinear optimization.  We provide details showing that the MLEs for $\delta$ and $\gamma$ maximize $P_{\mathbf{x}'}(M_c|\mathbf{y})$, which provides insight for interpretation of MLE recalibration across applications.  By maximizing the LLO likelihood, one maximizes $P_{\mathbf{x}'}(M_c|\mathbf{y})$.  We discuss various optimization routines considered for estimation of MLEs and boldness-recalibration parameters. While some additional intuition related to LLO-adjustment and boldness-recalibration is provided in this paper, future work includes additional mathematical rigor to show that boldness (measured by the standard deviation) increases with $\gamma$ and that the objective function in \eqref{objective_rj} is unimodal within the nonlinear constraint in \eqref{constraint_rj}.

This paper also presents the \textbf{BRcal} package. The \textbf{BRcal} package offers a variety of tools for assessing and improving probability predictions.  One key contribution of this R package is the \verb|brcal()| function which allows users to maximize the boldness (spread) in predictions while maintaining a specified level of calibration as measured by the posterior model probability of calibration.  The other main contributions of the \textbf{BRcal} package are (i) MLE recalibration, (ii) calibration assessment, and (iii) supporting visualizations. Lastly, we provide a new case study related to foreclosure predictions to further demonstrate the \textbf{BRcal} package and boldness-recalibration. Overall, the \textbf{BRcal} package is a useful tool to achieve more actionable predictions for better decision making. The \textbf{BRcal} package, with documentation and a user-friendly tutorial are available to users on CRAN and Github.



\backmatter

\vspace{1em}
\noindent

\noindent \textbf{Conflict of Interest} On behalf of all authors, the corresponding author states that there is no conflict of interest.

\begin{appendices}

\section{Derivation of Gradients}\label{app:gradients}

This appendix shows the derivation of the gradients shown in Sections \ref{app:mleobjgrad}, \ref{app:objgrad} and \ref{app:constrgrad} using related intermediate results presented in Section \ref{app:intres}.

\subsection{Intermediate Results} \label{app:intres}

 The intermediate results in this section are useful for more clearly and concisely writing the derivations to follow.  Let $x_i' = c(x_i; \delta, \gamma) = \frac{\delta x_i^\gamma}{\delta x_i^\gamma + (1-x_i)^\gamma}$ and $\hat{\delta} = \hat{\delta}_{MLE}'$ and $\hat{\gamma} = \hat{\gamma}_{MLE}'$ be the MLEs for $x_i'$.  We redefine these values for conciseness in derivations throughout the appendices. Note $x_{i,MLE}' = c(x_i'; \hat{\delta},\hat{\gamma})$, which is equivalent to $c(x_i; \hat{\delta} \delta^{\hat{\gamma}}, \gamma\hat{\gamma})$ by \ref{app:nestedLLO}. 

\subsubsection{Nested LLO-Adjustment} \label{app:nestedLLO}

Consider any LLO-adjusted prediction $x_i$ by $\delta$ and $\gamma$ that is then LLO-adjusted again by different parameters $\delta'$ and $\gamma'$.  The resulting prediction can be re-written as follows in terms of a single LLO-adjustment using both sets of parameters.
\begin{align*}
    c(c(x_i; \delta, \gamma); \delta', \gamma') &= \frac{\delta'(c(x_i; \delta, \gamma))^{\gamma'}}{\delta'(c(x_i; \delta, \gamma))^{\gamma'} + (1-c(x_i; \delta, \gamma))^{\gamma'}} \\
    &= \frac{\delta'\left(\frac{\delta x_i^\gamma}{\delta x_i^\gamma + (1-x_i)^\gamma}\right)^{\gamma'}}{\delta'\left(\frac{\delta x_i^\gamma}{\delta x_i^\gamma + (1-x_i)^\gamma}\right)^{\gamma'} + \left(1-\frac{\delta x_i^\gamma}{\delta x_i^\gamma + (1-x_i)^\gamma}\right)^{\gamma'}} \\
    &= \frac{\frac{\delta'\delta^{\gamma'}x_i^{\gamma\gamma'}}{(\delta x_i^\gamma + (1-x_i)^\gamma)^{\gamma'}}}{\frac{\delta'\delta^{\gamma'}x_i^{\gamma\gamma'}}{(\delta x_i^\gamma + (1-x_i)^\gamma)^{\gamma'}} + \frac{(1-x_i)^{\gamma \gamma'}}{(\delta x_i^\gamma + (1-x_i)^\gamma)^{\gamma'}}} \\
    &= \frac{\delta'\delta^{\gamma'}x_i^{\gamma\gamma'}}{\delta'\delta^{\gamma'}x_i^{\gamma\gamma'} + (1-x_i)^{\gamma \gamma'}} \\
    &= c(x_i; \delta' \delta^{\gamma '}, \gamma \gamma')
\end{align*}

\subsubsection{Relationship Between LLO-adjusted $x_i$ and $1-x_i$}

As mentioned in \citep{GuthrieFranck2024}, if LLO-adjusted predicted probability for an ``event'' is defined by $x_i' = c(x_i; \delta, \gamma)$, then the corresponding LLO-adjusted prediction for a ``non-event'' is defined by $1 - x_i' = 1- c(x_i; \delta, \gamma)$.  This can be re-written as the following LLO-adjustment. 

\begin{align*}
    1 - x_i' &= 1 - c(x_i; \delta, \gamma) \\
    &= 1 - \frac{\delta x_i^\gamma}{\delta x_i^\gamma + (1-x_i)^\gamma} \\
    &= \frac{(1-x_i)^\gamma}{\delta x_i^\gamma + (1-x_i)^\gamma} \\
    &= \frac{\frac{1}{\delta} (1-x_i)^\gamma}{\frac{1}{\delta} (1-x_i)^\gamma + x_i^\gamma} \\
    &= c(1-x_i; \frac{1}{\delta}, \gamma)
\end{align*}

\subsubsection{Log LLO-Adjustment}

The following results pertains to log LLO-adjusted predictions needed for calculations of the gradients in Sections \ref{app:objgrad} and \ref{app:constrgrad}. 

\begin{align*}
    log(x_i') &= log(\delta) + \gamma log(x_i) - log(\delta x_i^\gamma + (1-x_i)^\gamma) \\
    \\
    log(1 - x_i') &= log\left(c(1-x_i; \frac{1}{\delta}, \gamma)\right) \\
    &= \gamma log(1-x_i) -log\left((1-x_i)^\gamma + \delta x_i^\gamma \right) \\
    \\
    log(x_{i,MLE}') &= log( \hat{\delta} \delta^{\hat{\gamma}}) + \gamma\hat{\gamma}log(x_i) - log( \hat{\delta} \delta^{\hat{\gamma}} x_i^{\gamma\hat{\gamma}} + (1-x_i)^{\gamma\hat{\gamma}}) \\
    &= log( \hat{\delta}) + {\hat{\gamma}} log(\delta) + \gamma\hat{\gamma}log(x_i) - log( \hat{\delta} \delta^{\hat{\gamma}} x_i^{\gamma\hat{\gamma}} + (1-x_i)^{\gamma\hat{\gamma}}) \\
    \\
    log(1-x_{i,MLE}') &= \gamma\hat{\gamma} log(1-x_i) -log\left((1-x_i)^{\gamma\hat{\gamma}} + \hat{\delta} \delta^{\hat{\gamma}}x_i^{\gamma\hat{\gamma}} \right) \\
\end{align*}

\subsubsection{Intermediate Partial Derivatives}

The following results pertain to derivatives needed for calculations of the gradients in Sections \ref{app:objgrad} and \ref{app:constrgrad}. 

\begin{align*}
    \frac{\partial}{\partial\delta} x_i' &= \frac{(\delta x_i ^\gamma + (1-x_i)^\gamma)x_i^\gamma - \delta x_i^{2\gamma}}{(\delta x_i^\gamma + (1-x_i)^\gamma)^2}\\
    &= \frac{x_i^\gamma (1-x_i)^\gamma}{(\delta x_i^\gamma + (1-x_i)^\gamma)(\delta x_i^\gamma + (1-x_i)^\gamma)} \\
    &= \frac{1}{\delta} \frac{\delta x_i^\gamma}{\delta x_i^\gamma + (1-x_i)^\gamma} \frac{(1-x_i)^\gamma}{\delta x_i^\gamma + (1-x_i)^\gamma}\\
    &= \frac{1}{\delta}c(x_i; \delta, \gamma) c(1-x_i; \frac{1}{\delta}, \gamma) \\
    &= \frac{1}{\delta}x_i'(1-x_i')\\
    \\
    \frac{\partial}{\partial\gamma} x_i' &= \frac{(\delta x_i^\gamma + (1-x_i)^\gamma)\delta x_i^\gamma log(x_i) - \delta x_i^\gamma(\delta x_i^\gamma log(x_i) + (1-x_i)^\gamma log(1-x_i))}{(\delta x_i^\gamma + (1-x_i)^\gamma)^2}\\
    &= \frac{\delta x_i^\gamma (1-x_i)^\gamma }{(\delta x_i^\gamma + (1-x_i)^\gamma)(\delta x_i^\gamma + (1-x_i)^\gamma)} log\left(\frac{x_i}{1-x_i}\right)\\
    &= c(x_i; \delta, \gamma) c(1-x_i; \frac{1}{\delta}, \gamma) log\left(\frac{x_i}{1-x_i}\right)\\
    &= x_i'(1-x_i') log\left(\frac{x_i}{1-x_i}\right)\\
    \\
    \frac{\partial}{\partial\delta} log(x_i') &= \frac{1}{\delta} - \frac{x_i^\gamma}{\delta x_i^\gamma + (1-x_i)^\gamma}\\
    &= \frac{(1-x_i)^\gamma}{\delta\left(\delta x_i^\gamma + (1-x_i)^\gamma\right)}\\
    &= \frac{1}{\delta} c(1-x_i; \frac{1}{\delta}, \gamma) \\
    &= \frac{1}{\delta} (1 - x_i')\\
    &\\
    \frac{\partial}{\partial\gamma} log(x_i') &= log(x_i) - \frac{\delta x_i^\gamma log(x_i) + (1-x_i)^\gamma log(1-x_i)}{\delta x_i^\gamma + (1-x_i)^\gamma}\\
    &= \frac{(1-x_i)^\gamma log\left(\frac{x_i}{1-x_i}\right)}{\delta x_i^\gamma + (1-x_i)^\gamma} \\
    &= log\left(\frac{x_i}{1-x_i}\right)c(1-x_i; \frac{1}{\delta}, \gamma) \\
    &= log\left(\frac{x_i}{1-x_i}\right)(1 - x_i')
\\
    \frac{\partial}{\partial\delta} log(1-x_i') &= - \frac{x_i^\gamma}{\delta x_i^\gamma + (1-x_i)^\gamma} \\
    &= -\frac{1}{\delta} \frac{\delta x_i^\gamma}{\delta x_i^\gamma + (1-x_i)^\gamma} \\
    &= -\frac{1}{\delta} x_i'\\
    \\
    \frac{\partial}{\partial\gamma} log(1-x_i') &= log(1-x_i) - \frac{log(1-x_i)(1-x_i)^\gamma + log(x_i) \delta x_i^\gamma}{\delta x_i^\gamma + (1-x_i)^\gamma}\\
    &= log\left(\frac{1-x_i}{x_i}\right)\frac{\delta x_i^\gamma}{\delta x_i^\gamma + (1-x_i)^\gamma}\\
    &= log\left(\frac{1-x_i}{x_i}\right) x_i'\\
    &= -log\left(\frac{x_i}{1-x_i}\right) x_i'
\\
    \frac{\partial}{\partial\delta} log(x_{i,MLE}') &= \frac{\hat\gamma}{\delta} - \frac{\hat{\delta}\hat{\gamma} \delta^{\hat{\gamma}-1} x_i^{\gamma\hat{\gamma}}}{ \hat{\delta} \delta^{\hat{\gamma}} x_i^{\gamma\hat{\gamma}} + (1-x_i)^{\gamma\hat{\gamma}}}\\
    &= \frac{\hat\gamma (1-x_i)^{\gamma\hat{\gamma}}}{\delta(\hat{\delta} \delta^{\hat{\gamma}} x_i^{\gamma\hat{\gamma}} + (1-x_i)^{\gamma\hat{\gamma}})}\\
    &= \frac{\hat\gamma}{\delta}c(1-x_i; \frac{1}{\hat{\delta} \delta^{\hat{\gamma}}}, \gamma\hat{\gamma})\\
    &= \frac{\hat\gamma}{\delta} (1-x_{i,MLE}')\\
    \\
    \frac{\partial}{\partial\gamma} log(x_{i,MLE}') &= \hat\gamma log(x_i) - \frac{ \hat{\delta} \delta^{\hat{\gamma}} x_i^{\gamma\hat{\gamma}} log(x_i) \hat\gamma + (1-x_i)^{\gamma\hat{\gamma}}log(1-x_i)\hat\gamma}{ \hat{\delta} \delta^{\hat{\gamma}} x_i^{\gamma\hat{\gamma}} + (1-x_i)^{\gamma\hat{\gamma}}} \\
    &= \frac{\hat\gamma log(\frac{x_i}{1-x_i})(1-x_i)^{\gamma\hat\gamma}}{ \hat{\delta} \delta^{\hat{\gamma}} x_i^{\gamma\hat{\gamma}} + (1-x_i)^{\gamma\hat{\gamma}}} \\
    &= \hat\gamma log\left(\frac{x_i}{1-x_i}\right)c(1-x_i;\frac{1}{\hat{\delta} \delta^{\hat{\gamma}}}, \gamma\hat{\gamma})\\
    &= \hat\gamma log\left(\frac{x_i}{1-x_i}\right)  (1-x_{i,MLE}')\\
\\
    \frac{\partial}{\partial\delta} log(1-x_{i,MLE}') &= -\frac{\hat{\delta} \hat\gamma \delta^{\hat{\gamma}-1}x_i^{\gamma\hat{\gamma}}}{\hat{\delta} \delta^{\hat{\gamma}}x_i^{\gamma\hat{\gamma}} + (1-x_i)^{\gamma\hat{\gamma}}}\\
    &= -\frac{\hat\gamma}{\delta} \frac{\hat{\delta}  \delta^{\hat{\gamma}}x_i^{\gamma\hat{\gamma}}}{\hat{\delta} \delta^{\hat{\gamma}}x_i^{\gamma\hat{\gamma}} + (1-x_i)^{\gamma\hat{\gamma}}}\\
    &= -\frac{\hat\gamma}{\delta} x_{i,MLE}'\\
    \\
    \frac{\partial}{\partial\gamma} log(1-x_{i,MLE}') &=\hat\gamma log(1-x_i) - \frac{(1-x_i)^{\gamma\hat{\gamma}} log(1-x_i) \hat\gamma + \hat{\delta} \delta^{\hat{\gamma}}x_i^{\gamma\hat{\gamma}} log(x_i)\hat\gamma}{(1-x_i)^{\gamma\hat{\gamma}} + \hat{\delta} \delta^{\hat{\gamma}}x_i^{\gamma\hat{\gamma}}} \\
    &= \frac{\hat{\delta} \delta^{\hat{\gamma}}x_i^{\gamma\hat{\gamma}}\hat\gamma log(\frac{1-x_i}{x_i})}{\hat{\delta} \delta^{\hat{\gamma}}x_i^{\gamma\hat{\gamma}} +(1-x_i)^{\gamma \hat\gamma}}\\
    &= \hat\gamma log(\frac{1-x_i}{x_i}) c(x_i; \hat{\delta} \delta^{\hat{\gamma}}, \gamma \hat\gamma) \\
    &= \hat\gamma log(\frac{1-x_i}{x_i}) x_{i,MLE}'\\
    &= -\hat\gamma log(\frac{x_i}{1-x_i}) x_{i,MLE}'
\end{align*}

\subsection{MLE Objective Function Gradients}\label{app:mleobjgrad}

The following series of equations show the derivation of the gradients of the objective function in \eqref{eq:objmle} using the results from Section \ref{app:intres}.  We will derive these in terms of $\delta$ and then reparameterize in terms of $\tau$.

\begin{align*}
    \frac{\partial h(\delta, \gamma)}{ \partial \delta} &= -\sum_{i=1}^n y_i \frac{\partial}{\partial \delta} log(x_i') + (1-y_i)\frac{\partial}{\partial \gamma} log(1-x_i')\\
    &= -\sum_{i=1}^n y_i \frac{1}{\delta} (1 - x_i') + (1-y_i)(-\frac{1}{\delta} x_i')\\
    &= \sum_{i=1}^n \frac{1}{\delta} (x_i'(1-y_i) -y_i(1-x_i'))\\
    &= \sum_{i=1}^n \frac{1}{\delta} (x_i' - y_i)\\
    &= \sum_{i=1}^n \frac{1}{\delta} (c(x_i; \delta, \gamma) - y_i) \\
    \implies \frac{\partial h(\tau, \gamma)}{ \partial \tau} &= \sum_{i=1}^n \frac{1}{exp(\tau)} (c(x_i; exp(\tau), \gamma) - y_i) \frac{\partial}{\partial \tau} exp(\tau)\\
    &= \sum_{i=1}^n  (c(x_i; exp(\tau), \gamma) - y_i)
\end{align*}

\begin{align*}
    \frac{\partial h(\delta, \gamma)}{ \partial \gamma} &= -\sum_{i=1}^n y_i \frac{\partial}{\partial \gamma} log(x_i') + (1-y_i)\frac{\partial}{\partial \gamma} log(1-x_i')\\
    &= -\sum_{i=1}^n y_i log\left(\frac{x_i}{1-x_i}\right)(1-x_i') + (1-y_i)(-log\left(\frac{x_i}{1-x_i}\right)x_i')\\
    &= \sum_{i=1}^n log\left(\frac{x_i}{1-x_i}\right)(x_i'(1-y_i) -y_i(1-x_i'))\\
    &= \sum_{i=1}^n log\left(\frac{x_i}{1-x_i}\right) (x_i' - y_i)\\
    &= \sum_{i=1}^n log\left(\frac{x_i}{1-x_i}\right) (c(x_i; \delta, \gamma) - y_i) \\
    \implies     \frac{\partial h(\tau, \gamma)}{ \partial \gamma} &= \sum_{i=1}^n log\left(\frac{x_i}{1-x_i}\right) (c(x_i; exp(\tau), \gamma) - y_i)
\end{align*}

\subsection{Boldness-Recalibration Objective Function Gradients} \label{app:objgrad}

The following series of equations show the derivation of the gradients of the objective function in \eqref{objective_rj} using the results from Section \ref{app:intres}.

\begin{align*}
    \frac{\partial f(\delta, \gamma)}{ \partial \delta} &= -\frac{\frac{\partial}{ \partial \delta} \frac{1}{n-1} \sum_{i=1}^n (x_i' - \bar{x}')^2}{2 \sqrt{\frac{1}{n-1} \sum_{i=1}^n (x_i' - \bar{x}')^2}}\\
    &= -\frac{\sum_{i=1}^n (x_i' - \bar{x}')\frac{\partial}{ \partial \delta}(x_i' - \bar{x}')}{(n-1) sd(\mathbf{x'})}\\
    &= -\frac{\sum_{i=1}^n (x_i' - \bar{x}')\left(\frac{1}{\delta}x_i'(1-x_i') - \frac{1}{n} \sum_{j=1}^n \frac{1}{\delta}x_j'(1-x_j')\right)}{(n-1) sd(\mathbf{x'})}\\
        &= -\frac{\sum_{i=1}^n (x_i' - \bar{x}')\left(x_i'(1-x_i') - \frac{1}{n} \sum_{j=1}^n x_j'(1-x_j')\right)}{\delta(n-1) sd(\mathbf{x'})}\\
    \\
    \frac{\partial f(\delta, \gamma)}{ \partial \gamma} &= -\frac{\sum_{i=1}^n (x_i' - \bar{x}')\frac{\partial}{ \partial \gamma}(x_i' - \bar{x}')}{(n-1) sd(\mathbf{x'})}\\
    &= -\frac{\sum_{i=1}^n (x_i' - \bar{x}')\left(x_i'(1-x_i')log\left(\frac{x_i}{1-x_i}\right) - \frac{1}{n} \sum_{j=1}^n x_j'(1-x_j')log\left(\frac{x_j}{1-x_j}\right)\right)}{(n-1) sd(\mathbf{x'})}\\
\end{align*}

Thus, we define the Jacobian of (\ref{objective_rj})  as $J_f(\delta, \gamma) = \left[ \frac{\partial f}{\partial \delta}, \frac{\partial f}{\partial \gamma} \right]$  where
\begin{align} \label{jaq_objective_rj} 
    \frac{\partial f}{\partial \delta} &= \frac{-1}{\delta(n-1)sd(\mathbf{x}')} \sum_{i=1}^n  (x_i' - \bar{x}')\left(x_i'(1-x_i') - \frac{1}{n}\sum_{j=1}^n x_j'(1-x_j')\right),  \\
    \frac{\partial f}{\partial \gamma} &= \frac{-1}{(n-1)sd(\mathbf{x}')} \sum_{i=1}^n  (x_i' - \bar{x}')\left( log\left(\frac{x_i}{1-x_i}\right)x_i'(1-x_i') - \frac{1}{n}\sum_{j=1}^n log\left(\frac{x_j}{1-x_j}\right)x_j'(1-x_j')\right).
\end{align}

\subsection{Boldness-Recalibration Constraint Function Gradients} \label{app:constrgrad}

The following series of equations show the derivation of the gradients of the constraint function in  \eqref{constraint_rj} using the results from Section \ref{app:intres}.

\begin{align*}
    \frac{\partial g(\delta, \gamma)}{ \partial \delta} &= \frac{1}{\left(1+BF\frac{P(M_u)}{P(M_c)} \right)^2} \frac{\partial}{\partial \delta}\left(1+BF\frac{P(M_u)}{P(M_c)} \right)\\
    &= P_{\mathbf{x'}}(M_c|\mathbf{y})^2 \frac{P(M_u)}{P(M_c)} exp\left\{ -\frac{1}{2}(BIC_u - BIC_c)\right\} \frac{\partial}{\partial \delta} -\frac{1}{2}(BIC_u - BIC_c)\\
    &= P_{\mathbf{x'}}(M_c|\mathbf{y})^2 \left(\frac{1}{P_{\mathbf{x'}}(M_c|\mathbf{y})} -1\right) \\
    & \frac{\partial}{\partial \delta}\left(-log(n) + \sum_{i=1}^n y_i log(x_{i,MLE}') +(1-y_i)log(1-x_{i,MLE}') -y_i log(x_i') -(1-y_i)log(1-x_i') \right) \\
    &= P_{\mathbf{x'}}(M_c|\mathbf{y})(1-P_{\mathbf{x'}}(M_c|\mathbf{y})) \\
    & \left( \sum_{i=1}^n y_i \frac{\hat\gamma}{\delta} (1-x_{x_i,MLE}') + (1-y_i)(-\frac{\hat\gamma}{\delta}x_{i,MLE}') -y_i\frac{1}{\delta}(1-x_i')-(1-y_i)(-\frac{1}{\delta}x_i')\right)\\
    &= P_{\mathbf{x'}}(M_c|\mathbf{y})(1-P_{\mathbf{x'}}(M_c|\mathbf{y})) \left( \sum_{i=1}^n \frac{\hat\gamma}{\delta} y_i -\frac{\hat\gamma}{\delta} x_{i,MLE}' -\frac{1}{\delta} y_i +\frac{1}{\delta}x_i'\right)\\
    &= P_{\mathbf{x'}}(M_c|\mathbf{y})(1-P_{\mathbf{x'}}(M_c|\mathbf{y})) \frac{1}{\delta} \left( \sum_{i=1}^n (\hat\gamma -1)y_i - \hat\gamma x_{i,MLE}' +x_i'\right) \\
    \\
    \frac{\partial g(\delta, \gamma)}{ \partial \gamma}  &= P_{\mathbf{x'}}(M_c|\mathbf{y})(1-P_{\mathbf{x'}}(M_c|\mathbf{y})) \\
    &\frac{\partial}{\partial \gamma}\left(-log(n) + \sum_{i=1}^n y_i log(x_{i,MLE}') +(1-y_i)log(1-x_{i,MLE}') -y_i log(x_i') -(1-y_i)log(1-x_i') \right) \\
    &= P_{\mathbf{x'}}(M_c|\mathbf{y})(1-P_{\mathbf{x'}}(M_c|\mathbf{y})) \bigg( \sum_{i=1}^n \hat\gamma y_i log\left(\frac{x_i}{1-x_i}\right)(1-x_{i,MLE}')\\
    & - \hat\gamma (1-y_i) log\left(\frac{x_i}{1-x_i}\right)x_{i,MLE}'-y_i log\left(\frac{x_i}{1-x_i}\right)(1-x_i') + (1-y_i)  log\left(\frac{x_i}{1-x_i}\right) x_i' \bigg) \\
    &= P_{\mathbf{x'}}(M_c|\mathbf{y})(1-P_{\mathbf{x'}}(M_c|\mathbf{y}))  \bigg(\sum_{i=1}^n \hat\gamma y_i log\left(\frac{x_i}{1-x_i}\right)  -\hat\gamma log\left(\frac{x_i}{1-x_i}\right)  x_{i,MLE}' \\
    &- y_i log\left(\frac{x_i}{1-x_i}\right)  + log\left(\frac{x_i}{1-x_i}\right)  x_i' \bigg)\\
    &= P_{\mathbf{x'}}(M_c|\mathbf{y})(1-P_{\mathbf{x'}}(M_c|\mathbf{y}))\bigg(\sum_{i=1}^n log\left(\frac{x_i}{1-x_i}\right) \big((\hat\gamma-1)y_i - \hat\gamma x_{i,MLE}' +x_i'\big) \bigg)
\end{align*}

Thus, we define the Jacobian of (\ref{constraint_rj}) as $J_g(\delta, \gamma) = \left[ \frac{\partial g}{\partial \delta}, \frac{\partial g}{\partial \gamma} \right]$ where
\begin{align} \label{jaq_constraint_rj}
    \frac{\partial g}{\partial \delta} &= P_{\mathbf{x'}}(M_c|\mathbf{y})(1-P_{\mathbf{x'}}(M_c|\mathbf{y})) \frac{1}{\delta} \left( \sum_{i=1}^n \hat\gamma_{MLE}'(y_i - x_{i,MLE}') - y_i +x_i'\right),\\
    \frac{\partial g}{\partial \gamma} &= P_{\mathbf{x'}}(M_c|\mathbf{y})(1-P_{\mathbf{x'}}(M_c|\mathbf{y}))\bigg(\sum_{i=1}^n log\left(\frac{x_i}{1-x_i}\right) \big(\hat\gamma_{MLE}'(y_i - x_{i,MLE}') - y_i + x_i'\big) \bigg). 
\end{align}

\section{Strategy for Specifying Parameter Grid in \texttt{plot\_params()} } \label{app:plt_params}


When using \verb|plot_params()|, some care is required for the specifications of arguments \verb|dlim|, \verb|glim|, and \verb|k|, which together determine the fineness of your grid of parameter values plotted. 
We suggest the following scheme for determining the grid of parameter values to search. First, fix \verb|k| at some small number, less than 20 for sake of computation time. Then, center a grid with small range around $\hat{\delta}_{MLE}$ and $\hat{\gamma}_{MLE}$ for $\mathbf{x}$. A grid that is too course may not be able to detect regions of non-zero probability of calibration. Increase the grid fineness (i.e., increase \texttt{k}) until the probability of calibration near the MLEs on the plot approximately matches the probability of calibration from \verb|bayes_ms()| for the same set of data. At this point, you may find that your grid does not fully encompass the range of parameters values which produce recalibrated probabilities with meaningfully high posterior probability of being calibrated. If this is the case, expand your grid until the region with high probability of calibration is covered. Then, further increase \verb|k| to create a fine grid of values.  Additionally, we caution users from including $\gamma = 0$ in the grid.  This setting recalibrates all values in $\mathbf{x}$ to a single value which is not desirable in practice.  Unless the singular value is near the base rate, the set will be poorly calibrated and minimally bold, which does not align with the goal of boldness-recalibration.  

As mentioned, the computation time to construct this plot can be high if the grid of parameter values is very fine or the sample size of predictions is very large.  In an effort to reduce this bottleneck, we have developed a strategy to circumvent unnecessary calls to \verb|optim()|.  To calculate $P_{\mathbf{x}'_{ij}}(M_c|\mathbf{y})$ for each grid cell, this requires evaluating the likelihood at the MLEs for each LLO-adjusted set $\mathbf{x}'_{ij}$, which in turn requires a call to \verb|optim()| to get the MLEs.  However, given the MLEs for the original set $\mathbf{x}$ and the $\delta_i$ and $\gamma_j$ from the grid, we show that it is possible to get the MLEs for $\mathbf{x}'_{ij}$ without optimization via the result in Appendix \ref{app:pointslope}.  Thus, we can reduce the number of \verb|optim()| calls from one call for each grid cell to just one call overall to find the MLEs for the original set.  A direct result of this is that this reduces the computation time from 12.03 minutes to 32.88 second for the hockey data and from 1.03 hours to 2.26 minutes for the foreclosure data, on average for \texttt{k}=200 (on Ubuntu 22.04, R version 4.4.2).

\section{\texttt{plot\_params()} Speed Up} \label{app:pointslope}

In the plot produced by \verb|plot_params()|, recall each grid cell is color coded by 
\begin{align*}
    P_{\mathbf{x}'_{ij}}(M_c|\mathbf{y}) &= \left(1+\frac{P(M_u)}{P(M_c)} exp \{-\frac{1}{2}(BIC_u - BIC_c) \}\right)^{-1}
\end{align*}
where $BIC_u =  2\times log(n) - 2\times log(\pi(\hat\delta'_{MLE}, \hat\gamma'_{MLE}|\mathbf{x}_{ij}',\mathbf{y})))$, $\mathbf{x}_{ij}'=c(\mathbf{x};\delta_i,\gamma_j)$, and $\hat\delta_{MLE}'$ and $\hat\gamma_{MLE}'$ are the MLEs for $\mathbf{x}_{ij}'$.   Notice that $BIC_u$ requires a unique calculation of $\hat\delta_{MLE}'$ and $\hat\gamma_{MLE}'$ for each $i,j$. In this section, we show that $\hat\delta_{MLE}'$ and $\hat\gamma_{MLE}'$ can be calculated using only the MLEs for $\mathbf{x}$ ($\hat\delta_{MLE}$ and $\hat\gamma_{MLE}$), and the grid shift parameters ($\delta_i$ and $\gamma_j$).  This is important as it circumvents an additional call to \verb|optim()| for each grid cell.

First, note that
\begin{align} \label{app:MLEs}
    (\hat\delta_{MLE}, \hat\gamma_{MLE}) = argmax_{(\hat \delta,\hat \gamma)} \left(\prod_{k=1}^n c(x_k; \hat\delta, \hat\gamma)^{y_k} (1-c(x_k; \hat\delta, \hat\gamma))^{1-y_k}\right).
\end{align} 
Similarly,
\begin{align} \label{app:MLEprimes}
    (\hat\delta_{MLE}', \hat\gamma_{MLE}') &= argmax_{(\hat \delta',\hat \gamma')} \left(\prod_{k=1}^n c(x_k'; \hat\delta', \hat\gamma')^{y_k} (1-c(x_k'; \hat\delta', \hat\gamma'))^{1-y_k}\right)\\
    &= argmax_{(\hat \delta',\hat \gamma')} \left(\prod_{k=1}^n c(c(x_k;\delta_i,\gamma_j); \hat\delta', \hat\gamma')^{y_k} (1-c(c(x_k;\delta_i,\gamma_j); \hat\delta', \hat\gamma'))^{1-y_k}\right) \\
    &= argmax_{(\hat \delta',\hat \gamma')} \left(\prod_{k=1}^n c(x_k; \hat\delta' \delta_i^{ \hat\gamma'}, \hat\gamma' \gamma_j)^{y_k} (1-c(x_k; \hat\delta' \delta_i^{ \hat\gamma'}, \hat\gamma' \gamma_j))^{1-y_k}\right)
\end{align}
by the result in Appendix \ref{app:nestedLLO}. This implies that when both Equation (\ref{app:MLEs}) and (\ref{app:MLEprimes}) are maximized, $\hat\delta_{MLE} = \hat\delta_{MLE}' \delta_i^{ \hat\gamma_{MLE}'}$ and $\hat\gamma_{MLE}=\hat\gamma_{MLE}' \gamma_j$. Solving for $\hat\delta_{MLE}'$ and $\hat\gamma_{MLE}'$ results in 
\begin{align}  
    \hat\delta_{MLE}' &= \frac{\hat\delta_{MLE}}{\delta_i^{ \hat\gamma_{MLE} / \gamma_j }} \label{eq:deltamleprime} \\ 
    \hat\gamma_{MLE}' &= \frac{\hat\gamma_{MLE}}{\gamma_j}.  \label{eq:gammamleprime}
\end{align}
Using these expression, we circumvent the need to use \verb|optim()| to optimize Equation (\ref{app:MLEprimes}) to get $\hat\delta_{MLE}'$ and  $\hat\gamma_{MLE}'$ for each grid cell.  Note that the expressions for $\hat\gamma_{MLE}'$ is undefined when $\gamma_j = 0$, which is a valid grid cell value.  In this case, all probabilities are set to 0.5, regardless of $\delta_i$.  

\section{Proof of Result 1} \label{app:MLEmaximize}

In Section \ref{sec:backgound_rj} and \cite{GuthrieFranck2024}, we claim that by LLO-adjusting a set of probability predictions by the MLEs for $\delta$ and $\gamma$, their calibration will be maximized.  In other words, we claim that $$(\hat\delta_{MLE}, \hat\gamma_{MLE}) = argmax_{( \delta, \gamma)}P_{\mathbf{x'}}(M_c|\mathbf{y}).$$  To show this, first note that maximizing $P_{\mathbf{x'}}(M_c|\mathbf{y})$ and minimizing $\frac{1}{P_{\mathbf{x'}}(M_c|\mathbf{y})}$ achieve the same optimum values for $\delta$ and $\gamma$ as $P_{\mathbf{x'}}(M_c|\mathbf{y})$ can only take on values between 0 and 1.  Then, we have 
\begin{align}
    argmax_{( \delta, \gamma)}P_{\mathbf{x'}}(M_c|\mathbf{y}) &= argmin_{( \delta, \gamma)}\frac{1}{P_{\mathbf{x'}}(M_c|\mathbf{y})}\\
    &= argmin_{( \delta, \gamma)}\frac{P_{\mathbf{x'}}(\mathbf{y}|M_c)P(M_c) + P_{\mathbf{x'}}(\mathbf{y}|M_u)P(M_u)}{P_{\mathbf{x'}}(\mathbf{y}|M_c)P(M_c)}\\
    &= argmin_{( \delta, \gamma)} 1 + \frac{P_{\mathbf{x'}}(\mathbf{y}|M_u)P(M_u)}{P_{\mathbf{x'}}(\mathbf{y}|M_c)P(M_c)}\\
    &= argmin_{( \delta, \gamma)} \frac{P_{\mathbf{x'}}(\mathbf{y}|M_u)}{P_{\mathbf{x'}}(\mathbf{y}|M_c)}, \label{eq:mle_bfratio}
\end{align}
as $P(M_c)$ and $P(M_u)$ do not change with $\delta$ and $\gamma$.  Under the BIC approximation to $BF_\mathbf{x'}$,  (\ref{eq:mle_bfratio}) is equivalent to
\begin{align}
    argmax_{( \delta, \gamma)} BF_{\mathbf{x}'} &\approx argmax_{( \delta, \gamma)} exp\left\{-\frac{1}{2}\left(BIC_u - BIC_c\right) \right\} \\
    &=argmin_{( \delta, \gamma)} -\frac{1}{2}BIC_u + \frac{1}{2} BIC_c. \label{eq:diffBICprime}
\end{align}
Here, 
\begin{align}
    -\frac{1}{2}BIC_u &= -log(n) + log (\pi(\hat\delta_{MLE}', \hat\gamma_{MLE}'|\mathbf{x}', \mathbf{y}))\\
    &=\sum_{i=1}^n y_i log(c(x_i'; \hat\delta_{MLE}', \hat\gamma_{MLE}')) + (1-y_i) log(1-c(x_i'; \hat\delta_{MLE}', \hat\gamma_{MLE}'))
\end{align}
where $\hat\delta_{MLE}'$ and $\hat\gamma_{MLE}'$ are the MLEs maximizing $\pi(\mathbf{y}|\delta', \gamma')$.  Plugging the expressions for $\hat\delta_{MLE}'$ and $\hat\gamma_{MLE}'$ from (\ref{eq:deltamleprime}) and (\ref{eq:gammamleprime}), we have
\begin{align}
    \sum_{i=1}^n y_i log(c(x_i'; \frac{\hat\delta_{MLE}}{\delta^{ \hat\gamma_{MLE} / \gamma }}, \frac{\hat\gamma_{MLE}}{\gamma})) + (1-y_i) log(1-c(x_i'; \frac{\hat\delta_{MLE}}{\delta^{ \hat\gamma_{MLE} / \gamma }}, \frac{\hat\gamma_{MLE}}{\gamma})) \label{eq:BICuprime}
\end{align}
where 
\begin{align}
    c(x_i'; \frac{\hat\delta_{MLE}}{\delta^{ \hat\gamma_{MLE} / \gamma }}, \frac{\hat\gamma_{MLE}}{\gamma}) &= c(c(x_i;\delta, \gamma); \frac{\hat\delta_{MLE}}{\delta^{ \hat\gamma_{MLE} / \gamma }}, \frac{\hat\gamma_{MLE}}{\gamma})\\
    &= c(x_i; \frac{\hat\delta_{MLE}}{\delta^{ \hat\gamma_{MLE} / \gamma }} \delta^{ \hat\gamma_{MLE} / \gamma }, \frac{\hat\gamma_{MLE}}{\gamma}\gamma)\\
    &=c(x_i; \hat\delta_{MLE}, \hat\gamma_{MLE})
\end{align}
via the result in Appendix \ref{app:nestedLLO}. Substituting back into (\ref{eq:BICuprime}), we have
$$-\frac{1}{2}BIC_u = \sum_{i=1}^n y_i log(c(x_i; \hat\delta_{MLE}, \hat\gamma_{MLE})) + (1-y_i) log(1-c(x_i; \hat\delta_{MLE}, \hat\gamma_{MLE}))$$
which implies that $-\frac{1}{2}BIC_u $ does not change with $\delta$ and $\gamma$. Thus, (\ref{eq:diffBICprime}) is equivalent to
\begin{align}
    argmin_{( \delta, \gamma)} \frac{1}{2} BIC_c &= argmin_{( \delta, \gamma)} -log(\pi(1,1,|\mathbf{x}', \mathbf{y}))\\
    &= argmax_{( \delta, \gamma)} log(\pi(1,1,|\mathbf{x}', \mathbf{y}))\\
    &= argmax_{( \delta, \gamma)} -\frac{1}{2}BIC_u = \sum_{i=1}^n y_i log(c(x_i'; 1, 1)) + (1-y_i) log(1-c(x_i'; 1, 1))\\
    &= argmax_{( \delta, \gamma)} -\frac{1}{2}BIC_u = \sum_{i=1}^n y_i log(c(x_i; \delta, \gamma)) + (1-y_i) log(1-c(x_i; \delta, \gamma))\\
    &=  argmax_{( \delta, \gamma)} log(\pi(\mathbf{y}|\delta, \gamma))\\
    &= (\hat\delta_{MLE}, \hat\gamma_{MLE}).
\end{align}
Thus $(\hat\delta_{MLE}, \hat\gamma_{MLE}) = argmax_{( \delta, \gamma)}P_{\mathbf{x'}}(M_c|\mathbf{y}).$ This indicates that MLE- recalibration adjusts $\mathbf{x}$ such that their probability of calibration is maximized.

\section{Optimization Algorithm Comparisons} \label{app:algcomps}

In this section, we present results pertaining to the speed, convergence, and stability of the optimization routines considered for use in the \textbf{BRcal} package.  Throughout the section, we use the \texttt{hockey} and \texttt{foreclosure} data discussed in Section \ref{subsec:brcal_data}. All computation was done in R version 4.4.2 on a Linux machine with Ubuntu 22.04.

\subsection{Comparing Algorithms in \texttt{optim()}} \label{app:optimalgcomps}

As mentioned in Section \ref{subsec:mlestimation}, maximum likelihood estimation of $\delta$ and $\gamma$ requires a numerical optimization routine.  In the \textbf{BRcal} package, we rely on the \texttt{optim()} function for this optimization.  Of the algorithms available in \texttt{optim()}, we consider Broyden–Fletcher–Goldfarb–Shanno (BFGS) algorithm, the conjugate gradients (CG) method, L-BFGS \citep{NocedalWright1998}, and Nelder-Mead \citep{nelder-mead}.  Additionally, each of these algorithms except Nelder-Mead optionally allow users to specify a closed-form gradient function rather than use a numerical approximation. From Equation \ref{eq:mle_minim}, our objective function to get the MLEs is

\begin{equation} \label{eq:objmle}
    h(\tau, \gamma) = -\sum_{i=1}^n y_i log(c(x_i; exp(\tau), \gamma)) + (1-y_i) log(1-c(x_i; exp(\tau), \gamma)).
\end{equation}
The Jacobian of Equation \eqref{eq:objmle}, which is derived in Section \ref{app:mleobjgrad}, is defined as $J_h(\tau, \gamma) = \left[\frac{\partial h}{\partial \delta}, \frac{\partial h}{\partial \gamma} \right]$, where 
\begin{align}
    \frac{\partial h(\tau, \gamma)}{ \partial \tau} &= \sum_{i=1}^n \frac{1}{exp(\tau)} (c(x_i; exp(\tau), \gamma) - y_i)\\
    \frac{\partial h(\tau, \gamma)}{ \partial \gamma} &= \sum_{i=1}^n log\left(\frac{x_i}{1-x_i}\right) (c(x_i; exp(\tau), \gamma) - y_i)
\end{align}

To gauge the sensitivity to the optimization starting location, we ran each algorithm over a 25 by 25 grid of starting values for $\delta$ and $\gamma$, where $\delta \in (0.05, 20)$ and $\gamma \in (-1, 10)$. We also compare the performance of the optimization with the original parameterization in terms of $\delta$ compared to the reparameterization in terms of $\tau$. In total, there were 87,500 runs under the described settings on our two datasets.

\begin{figure}[h!]
\begin{center}
\includegraphics[width=5in]{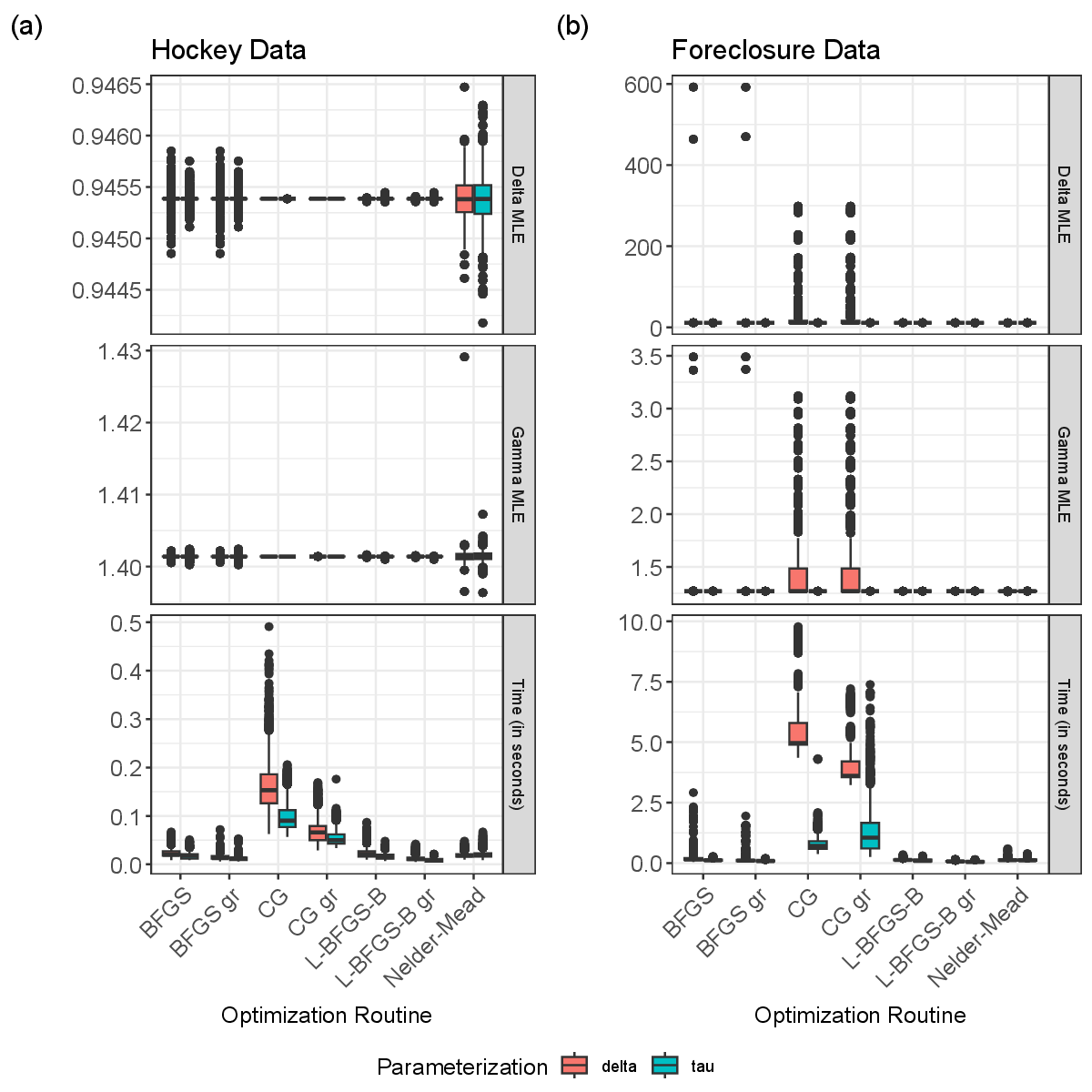}
\end{center}
\caption{Boxplots summarizing performance of optimization routines for the (a) \texttt{hockey} data and (b) \texttt{foreclosure} data. Optimization routines are shown on the x-axis.  Boxplot color represents whether the parameterization in terms of $\delta$ or $\tau$ was used. The first and second rows show the achieved MLES for $\delta$ and $\gamma$, respectively.  The last row shows the time (in seconds) each algorithm took to run.}
\label{fig:optim_all}
\end{figure}

Figure \ref{fig:optim_all} presents the achieved MLEs for $\delta$ and $\gamma$ in the first and second rows, respectively, and the speed per run (in seconds) in the third row, across algorithms. Table \ref{tab:optim_algs} further summarizes the average speed (in seconds), and variability in achieved MLEs for each algorithm. 

\begin{table}[]
\begin{tabular}{llcrrrr}
\hline
\textbf{Data}        & \textbf{Algorithm}   & \textbf{Param.} & \textbf{Avg. Time}  & \textbf{sd} $\mathbf{\hat \delta_{MLE}}$ & \textbf{sd} $\mathbf{\hat \gamma_{MLE}}$  \\
\hline\hline
\textbf{Hockey} & BFGS  & $\delta$  & 0.023 & 0.0001 & 0.0001 \\
            &             & $\tau$  & 0.018 & 0.0000 & 0.0002 \\
            & BFGS gr     & $\delta$  & 0.015 & 0.0001 & 0.0001 \\
            &             & $\tau$  & 0.012 & 0.0000 & 0.0002 \\
            & CG          & $\delta$  & 0.163 & 0.0000 & 0.0000 \\
            &             & $\tau$  & 0.100 & 0.0000 & 0.0000 \\
            & CG gr       & $\delta$  & 0.069 & 0.0000 & 0.0000 \\
            &             & $\tau$  & 0.056 & 0.0000 & 0.0000 \\
            & L-BFGS      & $\delta$  & 0.023 & 0.0000 & 0.0000\\
            &             & $\tau$  & 0.017 & 0.0000 & 0.0000 \\
            & L-BFGS gr   & $\delta$  & 0.012 & 0.0000 & 0.0000\\
            &             & $\tau$  & 0.009 & 0.0000 & 0.0000\\
            & Nelder-Mead & $\delta$  & 0.019 & 0.0002 & 0.0012 \\
            &             & $\tau$  & 0.020 & 0.0003 & 0.0007\\
\hline
\textbf{Foreclosure} & BFGS  & $\delta$  & 0.190  & 29.4174 &  0.1219 \\
            &             & $\tau$  & 0.124 & 0.0010 & 0.0001 \\
            & BFGS gr     & $\delta$  & 0.125 & 29.5697 & 0.1221 \\
            &             & $\tau$  & 0.090 & 0.0010 & 0.0001 \\
            & CG          & $\delta$  & 5.585 & 51.4699 & 0.4816 \\
            &             & $\tau$  & 0.782 & 0.0000 & 0.0000 \\
            & CG gr       & $\delta$  & 4.047 & 51.4103 & 0.4814 \\
            &             & $\tau$  & 1.314 & 0.0000 & 0.0000 \\
            & L-BFGS      & $\delta$  & 0.137 & 0.0002 & 0.0000 \\
            &             & $\tau$  & 0.106 & 0.0001 & 0.0000 \\
            & L-BFGS gr   & $\delta$  & 0.073 & 0.0001 & 0.0000 \\
            &             & $\tau$  & 0.055 & 0.0001 & 0.0000 \\
            & Nelder-Mead & $\delta$  & 0.132 & 0.0073 & 0.0005 \\
            &             & $\tau$  & 0.128 & 0.0054 & 0.0004 \\          
\hline
\end{tabular}
\caption{Table summarizing performance of optimization routines in \texttt{optim()} in terms of the average speed and variability in achieved MLEs for $\delta$ and $\gamma$ for the \texttt{hockey} and \texttt{foreclosure} datasets.}
\label{tab:optim_algs}
\end{table}

In general, we found optimizing with the $\tau$ reparameterization was faster and promoted improved convergence.  On average, optimizing in terms of $\tau$ was 1.37 times faster for the hockey data and 3.96 times faster for the foreclosure data compared to optimizing in terms of $\delta$.  When optimizing in terms of $\delta$, we found convergence was much less consistent for BFGS and CG, which is seen in that large standard deviations in Table \ref{tab:nlopt_algs} for $\hat \delta_{MLE}$ for the foreclosure data. 


\subsection{Comparing Algorithms in \texttt{nloptr()}}\label{app:nloptralgcomps}

To compare algorithms in \texttt{nloptr()}, we ran 95\% boldness-recalibration on our two datasets under a variety of settings. We first narrowed down the algorithms available from \textbf{NLopt} based on which accommodate both a nonlinear objective and a nonlinear inequality constraint.  Then, to further narrow down the options, we tried each of the possible algorithms on both datasets and eliminated those that did not run properly.  After this, we considered three algorithms: AUGLAG \citep{auglag1,auglag2}, MMA \citep{MMA}, and SLSQP \citep{slsqp1,slsqp2}.  AUGLAG had an option to use gradient information (LD) and an option to be derivative free (LN).  Additionally, since AUGLAG is a two-stage optimization, we considered COBYLA \citep{cobyla}, SLSQP, and MMA as the inner algorithms.  MMA also uses an inner algorithm, and for this we considered AUGLAGLD and MMA.  Lastly, we also included SLSQP on its own.  For each of these 9 algorithms, we also ran each under the $\delta$ parameterization of the objective function and the $\tau$ reparameterization. Since we suggest starting the optimization at the MLEs, we did not explore different starting values in this study.  To gauge the average time for each algorithm to run, we ran 50 replication of each algorithm and parameterization combination for both datasets.  As these are not stochastic approaches, all algorithms achieved the same values of the 95\% boldness-recalibration parameters and $s_b$.

\begin{figure}[h!]
\begin{center}
\includegraphics[width=5in]{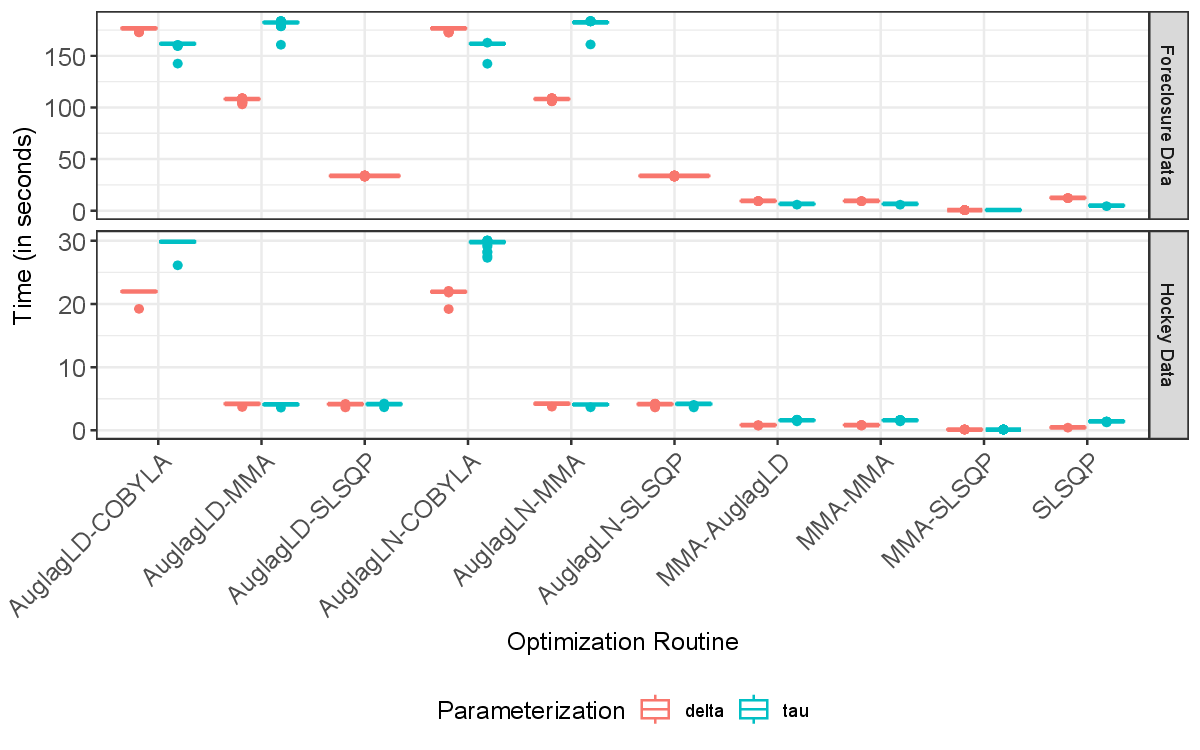}
\end{center}
\caption{Boxplots summarizing the speed of optimization routines for 95\% boldness-recalibration for the \texttt{hockey} data (top panel) and \texttt{foreclosure} data (bottom panel). Optimization routines are shown on the x-axis.  Boxplot color represents whether the parameterization in terms of $\delta$ or $\tau$ was used.}
\label{fig:nloptr_time}
\end{figure}

Table \ref{tab:nlopt_algs} summarizes the average time (in seconds), and the achieved values of $\hat \delta_{0.95}$, $\hat \gamma_{0.95}$, and $s_b$. Figure \ref{fig:nloptr_time} presents boxplots of the speed per run (in seconds) across algorithms between the $\delta$ and $\tau$ parameterizations. 

\begin{table}[] 
\begin{tabular}{llcrrrr}
\hline
\textbf{Data}        & \textbf{Algorithm}   & \textbf{Param.} & \textbf{Avg. Time} & $\mathbf{\hat \delta_{0.95}}$ & $\mathbf{\hat \gamma_{0.95}}$ & \textbf{Achieved $\mathbf{s_b}$}\\
\hline\hline
\textbf{Hockey}      & AuglagLN-COBYLA & $\delta$  & 21.885 & 0.873 & 1.959 & 0.165 \\
            &             & $\tau$  & 29.643 & 0.945 & 1.401 & 0.124 \\
            & AuglagLN-SLSQP & $\delta$  & 4.142 & 0.873 & 1.959 & 0.165 \\
            &             & $\tau$  & 4.168 & 0.873 & 1.959 & 0.165 \\
            & AuglagLN-MMA & $\delta$  & 4.212 & 0.873 & 1.959 & 0.165 \\
            &             & $\tau$  & 4.077 & 0.873 & 1.959 & 0.165 \\
            & AuglagLD-COBYLA & $\delta$  & 21.936 & 0.873 & 1.959 & 0.165 \\
            &             & $\tau$  & 29.777 & 0.945 & 1.401 & 0.124 \\
            & AuglagLD-SLSQP & $\delta$  & 4.143 & 0.873 & 1.959 & 0.165 \\
            &             & $\tau$  & 4.154 & 0.873 & 1.959 & 0.165 \\
            & AuglagLD-MMA & $\delta$  & 4.205 & 0.873 & 1.959 & 0.165 \\
            &             & $\tau$  & 4.081 & 0.873 & 1.959 & 0.165 \\
            & MMA-AuglagLD & $\delta$  & 0.833 & 0.873 & 1.959 & 0.165 \\
            &             & $\tau$  & 1.604 & 0.873 & 1.959 & 0.165 \\
            & MMA-MMA & $\delta$  & 0.833 & 0.873 & 1.959 & 0.165 \\
            &             & $\tau$  & 1.598 & 0.873 & 1.959 & 0.165 \\
            & SLSQP & $\delta$  & 0.469 & 0.873 & 1.959 & 0.165 \\
            &             & $\tau$  & 1.413 & 0.873 & 1.959 & 0.165 \\
            \hline
\textbf{Foreclosure}      & AuglagLN-COBYLA & $\delta$  & 176.358 & 11.109 & 1.271 & 0.179 \\
            &             & $\tau$  & 161.481 & 9.825 & 1.278 & 0.188 \\
            & AuglagLN-SLSQP & $\delta$  & 33.737 & 12.135 & 1.412 & 0.199\\
            &             & $\tau$  &   ---   &  ---  & --- & --- \\
            & AuglagLN-MMA & $\delta$  & 108.126 & 12.105 & 1.411 & 0.199 \\
            &             & $\tau$  & 182.174 & 11.109 & 1.271 & 0.179 \\
            & AuglagLD-COBYLA & $\delta$  & 176.458 & 11.109 & 1.271 & 0.179 \\
            &             & $\tau$  & 161.327 & 9.825 & 1.278 & 0.188 \\
            & AuglagLD-SLSQP & $\delta$  & 33.746 & 12.135 & 1.412 & 0.199\\
            &             & $\tau$  & ---         &      ---                  &        ---                &        ---      \\
            & AuglagLD-MMA & $\delta$  & 107.905 & 12.105 & 1.411 & 0.199 \\
            &             & $\tau$  & 181.811 & 11.109 & 1.271 & 0.179 \\
            & MMA-AuglagLD & $\delta$  & 9.437 & 12.131 & 1.412 & 0.199 \\
            &             & $\tau$  & 6.583 & 11.106 & 1.370 & 0.197 \\
            & MMA-MMA& $\delta$  & 9.436 & 12.131 & 1.412 &  0.199 \\
            &             & $\tau$  & 6.571 & 11.106 & 1.370 & 0.197 \\
            & SLSQP & $\delta$  & 12.466 & 12.131 & 1.412 & 0.199 \\
            &             & $\tau$  & 4.958 & 11.109 & 1.271  & 0.179\\
\hline
\end{tabular} 
\caption{Table summarizing performance of \texttt{nloptr()} routines for getting 95\% boldness-recalibration parameter estimates for $\delta$ and $\gamma$ in terms of average speed, achieved estimates, and achieved boldness for the \texttt{hockey} and \texttt{foreclosure} datasets.}
\label{tab:nlopt_algs}
\end{table}

\section{Additional Intuition Behind Boldness-Recalibration} \label{app:BRintuition}

In Figures \ref{fig:hockeycontours_ch3} and  \ref{fig:contourplot_rj}, we present color contour plots of the posterior model probability surface with points indicating where the 95, 90, and 80\% boldness-recalibration parameters lie on that surface.  We briefly discuss the intuition behind why the boldness-recalibration parameters tend to fall along their corresponding contours where values of $\gamma$ are larger in Section \ref{subsubsec:colorcontourplot}.  To further demonstrate this, Figure \ref{fig:sd_surfaces} below shows the negative objective surface ($s_b$) for the \texttt{hockey} data (left) and \texttt{foreclosure} data (right) with contours drawn of the constraint surface, $P_\mathbf{x'}(M_c|\mathbf{y}) \geq t$, for $t=0.1, 0.2, 0.3, 0.4, 0.5, 0.6, 0.7, 0.8,$ and $0.9$.  Boldness-recalibration parameters for all levels of $t$ (except 0.1) are plotted with white points.  Notice in both plots that within a particular constrained region, we see one unique maximum for $s_b$ at reasonable levels of calibration.  At low levels of calibration, where the posterior model probability surface begins to flatten out, optimization tends to be more unstable, which is why the parameters for $t=0.1$ are not shown. 

Notice that in both plots, boldness (as measured by $s_b$) increases with values of $\gamma$.  This is due to $\gamma$ directly adjusting the spread of predictions on the log odds scale, which translates to a similar, but lessened, change in spread on the probability scale when probabilities are approximately centered at 0.5.  When predictions are skewed toward 0 or 1, the increase of spread from $\gamma$ tails off once $\gamma$ becomes too large. ``Too large'' depends on sample size and the degree to which the probabilities are skewed.  However, we have found that ``too large'' tends to fall outside of the constrained region of interest.  In all cases explored, spread increases with $\gamma$ and thus the boldness-recalibration parameter estimates fall on along the upper boundary of the constraint. The location along the upper boundary also depends on $\delta$. In the left panel, boldness does not change with $\delta$, which is likely due to the fact that the \texttt{hockey} data did not require much shifting to achieve calibration, thus our constrained region is located where $\delta$ is close to 1 and little shifting is applied. Then, because spread increases with $\gamma$, we see the boldness-recalibration parameters for all levels of $t$, even at low levels of calibration, fall along the upper edge of their corresponding contour. In the right hand panel, we see boldness decreases with $\delta$, because the \texttt{foreclosure} data needed substantial shifting to achieve calibration, so our constrained region is located where $\delta$ is large and the substantial shifting affects how much spread the data can achieve.  We also see that for $t\geq0.4$, the boldness-recalibration estimate for $\gamma$ increases as $t$ decreases.  However, the estimates for $t=0.2$ and $0.3$ for $\gamma$ are lower than the higher levels of calibration due to the decrease in boldness with $\delta$.

\begin{figure}[h!]
\begin{center}
\includegraphics[width=5in]{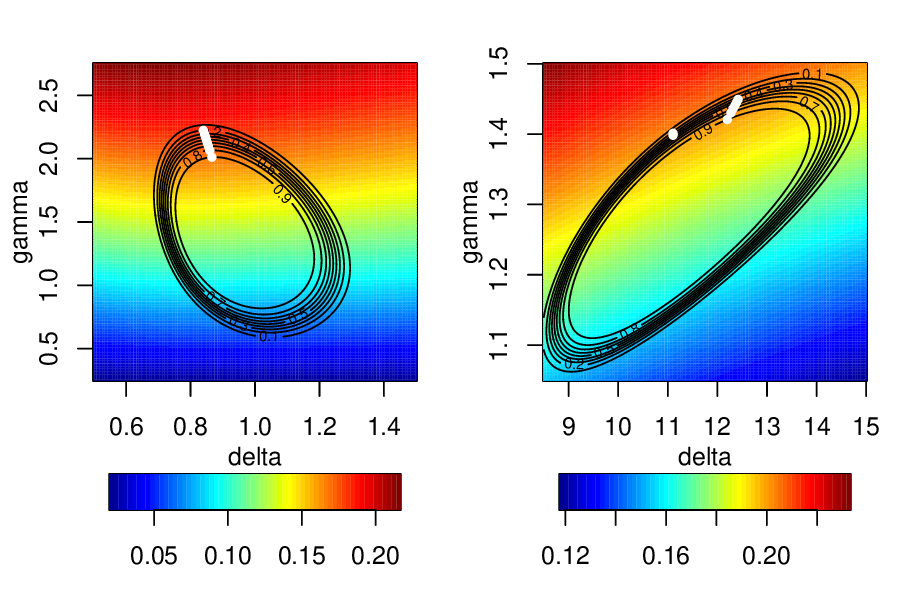}
\end{center}
\caption{Color contour plots summarizing the negative objective surface (standard deviation), using color, and the constraint surface (posterior probability of calibration), using contours, for boldness-recalibration for the \texttt{hockey} data (left) and \texttt{foreclosure} data (right).  Grid cell color represents the value of $s_b$ and the contours show constraints at $t=0.1, 0.2, 0.3, 0.4, 0.5, 0.6, 0.7, 0.8,$ and $0.9$.  The white points are boldness-recalibration estimates under each value of $t$ (except 0.1).}
\label{fig:sd_surfaces}
\end{figure}




\end{appendices} 


\bibliography{main}


\begin{thebibliography}{39}
\ifx \bisbn   \undefined \def \bisbn  #1{ISBN #1}\fi
\ifx \binits  \undefined \def \binits#1{#1}\fi
\ifx \bauthor  \undefined \def \bauthor#1{#1}\fi
\ifx \batitle  \undefined \def \batitle#1{#1}\fi
\ifx \bjtitle  \undefined \def \bjtitle#1{#1}\fi
\ifx \bvolume  \undefined \def \bvolume#1{\textbf{#1}}\fi
\ifx \byear  \undefined \def \byear#1{#1}\fi
\ifx \bissue  \undefined \def \bissue#1{#1}\fi
\ifx \bfpage  \undefined \def \bfpage#1{#1}\fi
\ifx \blpage  \undefined \def \blpage #1{#1}\fi
\ifx \burl  \undefined \def \burl#1{\textsf{#1}}\fi
\ifx \doiurl  \undefined \def \doiurl#1{\url{https://doi.org/#1}}\fi
\ifx \betal  \undefined \def \betal{\textit{et al.}}\fi
\ifx \binstitute  \undefined \def \binstitute#1{#1}\fi
\ifx \binstitutionaled  \undefined \def \binstitutionaled#1{#1}\fi
\ifx \bctitle  \undefined \def \bctitle#1{#1}\fi
\ifx \beditor  \undefined \def \beditor#1{#1}\fi
\ifx \bpublisher  \undefined \def \bpublisher#1{#1}\fi
\ifx \bbtitle  \undefined \def \bbtitle#1{#1}\fi
\ifx \bedition  \undefined \def \bedition#1{#1}\fi
\ifx \bseriesno  \undefined \def \bseriesno#1{#1}\fi
\ifx \blocation  \undefined \def \blocation#1{#1}\fi
\ifx \bsertitle  \undefined \def \bsertitle#1{#1}\fi
\ifx \bsnm \undefined \def \bsnm#1{#1}\fi
\ifx \bsuffix \undefined \def \bsuffix#1{#1}\fi
\ifx \bparticle \undefined \def \bparticle#1{#1}\fi
\ifx \barticle \undefined \def \barticle#1{#1}\fi
\bibcommenthead
\ifx \bconfdate \undefined \def \bconfdate #1{#1}\fi
\ifx \botherref \undefined \def \botherref #1{#1}\fi
\ifx \url \undefined \def \url#1{\textsf{#1}}\fi
\ifx \bchapter \undefined \def \bchapter#1{#1}\fi
\ifx \bbook \undefined \def \bbook#1{#1}\fi
\ifx \bcomment \undefined \def \bcomment#1{#1}\fi
\ifx \oauthor \undefined \def \oauthor#1{#1}\fi
\ifx \citeauthoryear \undefined \def \citeauthoryear#1{#1}\fi
\ifx \endbibitem  \undefined \def \endbibitem {}\fi
\ifx \bconflocation  \undefined \def \bconflocation#1{#1}\fi
\ifx \arxivurl  \undefined \def \arxivurl#1{\textsf{#1}}\fi
\csname PreBibitemsHook\endcsname

\bibitem[\protect\citeauthoryear{Guthrie and Franck}{2024}]{GuthrieFranck2024}
\begin{botherref}
\oauthor{\bsnm{Guthrie}, \binits{A.P.}},
\oauthor{\bsnm{Franck}, \binits{C.T.}}:
Boldness-recalibration for binary event predictions.
The American Statistician,
1--11
(2024)
\doiurl{10.1080/00031305.2024.2339266}
{\href{https://arxiv.org/abs/https://doi.org/10.1080/00031305.2024.2339266}{{https://doi.org/10.1080/00031305.2024.2339266}}}
\end{botherref}
\endbibitem

\bibitem[\protect\citeauthoryear{{De Cock Campo}}{2023}]{CalibrationCurves-package}
\begin{botherref}
\oauthor{\bsnm{{De Cock Campo}}, \binits{B.}}:
Towards reliable predictive analytics: a generalized calibration framework.
arXiv,
2309--08559
(2023)
\end{botherref}
\endbibitem

\bibitem[\protect\citeauthoryear{Rhodes}{2023}]{pmcalibration-package}
\begin{botherref}
\oauthor{\bsnm{Rhodes}, \binits{S.}}:
Pmcalibration: Calibration Curves for Clinical Prediction Models.
(2023).
R package version 0.1.0.
\url{https://github.com/stephenrho/pmcalibration}
\end{botherref}
\endbibitem

\bibitem[\protect\citeauthoryear{Dalton}{2013}]{Dalton2013}
\begin{barticle}
\bauthor{\bsnm{Dalton}, \binits{J.E.}}:
\batitle{Flexible recalibration of binary clinical prediction models}.
\bjtitle{Statistics in Medicine}
\bvolume{32}(\bissue{2}),
\bfpage{282}--\blpage{289}
(\byear{2013})
\doiurl{10.1002/sim.5544}
{\href{https://arxiv.org/abs/https://onlinelibrary.wiley.com/doi/pdf/10.1002/sim.5544}{{https://onlinelibrary.wiley.com/doi/pdf/10.1002/sim.5544}}}
\end{barticle}
\endbibitem

\bibitem[\protect\citeauthoryear{Zadrozny and Elkan}{2002}]{ZadroznyElkan2002}
\begin{botherref}
\oauthor{\bsnm{Zadrozny}, \binits{B.}},
\oauthor{\bsnm{Elkan}, \binits{C.P.}}:
Transforming classifier scores into accurate multiclass probability estimates.
Proceedings of the eighth ACM SIGKDD international conference on Knowledge discovery and data mining
(2002)
\end{botherref}
\endbibitem

\bibitem[\protect\citeauthoryear{Evans and Murphy}{2018}]{rfUtilities-package}
\begin{botherref}
\oauthor{\bsnm{Evans}, \binits{J.S.}},
\oauthor{\bsnm{Murphy}, \binits{M.A.}}:
rfUtilities.
(2018).
R package version 2.1-3.
\url{https://cran.r-project.org/package=rfUtilities}
\end{botherref}
\endbibitem

\bibitem[\protect\citeauthoryear{Klambauer and Mayr}{2015}]{platt-package}
\begin{botherref}
\oauthor{\bsnm{Klambauer}, \binits{G.}},
\oauthor{\bsnm{Mayr}, \binits{A.}}:
Platt: The Package Does Platt Scaling.
(2015).
R package version 0.99.4
\end{botherref}
\endbibitem

\bibitem[\protect\citeauthoryear{Platt}{2000}]{Platt2000}
\begin{botherref}
\oauthor{\bsnm{Platt}, \binits{J.}}:
Probabilistic outputs for support vector machines and comparisons to regularized likelihood methods.
Adv. Large Margin Classif.
\textbf{10}
(2000)
\end{botherref}
\endbibitem

\bibitem[\protect\citeauthoryear{Filho and Kull}{2017}]{betacal-package}
\begin{botherref}
\oauthor{\bsnm{Filho}, \binits{T.M.S.}},
\oauthor{\bsnm{Kull}, \binits{M.}}:
Betacal: Beta Calibration.
(2017).
R package version 0.1.0.
\url{https://CRAN.R-project.org/package=betacal}
\end{botherref}
\endbibitem

\bibitem[\protect\citeauthoryear{Kull et~al.}{2017}]{Kulletal2017}
\begin{bchapter}
\bauthor{\bsnm{Kull}, \binits{M.}},
\bauthor{\bsnm{Silva~Filho}, \binits{T.}},
\bauthor{\bsnm{Flach}, \binits{P.}}:
\bctitle{Beta calibration: a well-founded and easily implemented improvement on logistic calibration for binary classifiers}.
In: \bbtitle{Proceedings of the 20th International Conference on Artificial Intelligence and Statistics},
pp. \bfpage{623}--\blpage{631}
(\byear{2017})
\end{bchapter}
\endbibitem

\bibitem[\protect\citeauthoryear{Satop\"{a}\"{a}}{2022}]{Satopaa2022}
\begin{barticle}
\bauthor{\bsnm{Satop\"{a}\"{a}}, \binits{V.A.}}:
\batitle{Regularized aggregation of one-off probability predictions}.
\bjtitle{Oper. Res.}
\bvolume{70}(\bissue{6}),
\bfpage{3558}--\blpage{3580}
(\byear{2022})
\doiurl{10.1287/opre.2021.2224}
\end{barticle}
\endbibitem

\bibitem[\protect\citeauthoryear{Merkle and Steyvers}{2013}]{scoring-package}
\begin{barticle}
\bauthor{\bsnm{Merkle}, \binits{E.C.}},
\bauthor{\bsnm{Steyvers}, \binits{M.}}:
\batitle{Choosing a strictly proper scoring rule}.
\bjtitle{Decision Analysis}
\bvolume{10},
\bfpage{292}--\blpage{304}
(\byear{2013})
\end{barticle}
\endbibitem

\bibitem[\protect\citeauthoryear{Bosse et~al.}{2022}]{scoringutils-package}
\begin{barticle}
\bauthor{\bsnm{Bosse}, \binits{N.I.}},
\bauthor{\bsnm{Gruson}, \binits{H.}},
\bauthor{\bsnm{Cori}, \binits{A.}},
\bauthor{\bsnm{{van Leeuwen}}, \binits{E.}},
\bauthor{\bsnm{Funk}, \binits{S.}},
\bauthor{\bsnm{Abbott}, \binits{S.}}:
\batitle{Evaluating forecasts with scoringutils in r}.
\bjtitle{arXiv}
(\byear{2022})
\doiurl{10.48550/ARXIV.2205.07090}
\end{barticle}
\endbibitem

\bibitem[\protect\citeauthoryear{Signorell}{2024}]{DescTools-package}
\begin{botherref}
\oauthor{\bsnm{Signorell}, \binits{A.}}:
DescTools: Tools for Descriptive Statistics.
(2024).
R package version 0.99.53.
\url{https://CRAN.R-project.org/package=DescTools}
\end{botherref}
\endbibitem

\bibitem[\protect\citeauthoryear{Kuhn et~al.}{2024}]{yardstick-package}
\begin{botherref}
\oauthor{\bsnm{Kuhn}, \binits{M.}},
\oauthor{\bsnm{Vaughan}, \binits{D.}},
\oauthor{\bsnm{Hvitfeldt}, \binits{E.}}:
Yardstick: Tidy Characterizations of Model Performance.
(2024).
R package version 1.3.0.
\url{https://CRAN.R-project.org/package=yardstick}
\end{botherref}
\endbibitem

\bibitem[\protect\citeauthoryear{Brier}{1950}]{Brier1950}
\begin{barticle}
\bauthor{\bsnm{Brier}, \binits{G.W.}}:
\batitle{Verification of forecasts expressed in terms of probability}.
\bjtitle{Monthly Weather Review}
\bvolume{78}(\bissue{1}),
\bfpage{1}--\blpage{3}
(\byear{1950})
\doiurl{10.1175/1520-0493(1950)078<0001:VOFEIT>2.0.CO;2}
\end{barticle}
\endbibitem

\bibitem[\protect\citeauthoryear{Liu}{2024}]{calibration-package}
\begin{botherref}
\oauthor{\bsnm{Liu}, \binits{H.}}:
Calibration: Calibration Performance Evaluation and Recalibration.
(2024).
R package version 0.0.0.9000
\end{botherref}
\endbibitem

\bibitem[\protect\citeauthoryear{Matloff}{2021}]{calibtools-package}
\begin{botherref}
\oauthor{\bsnm{Matloff}, \binits{N.}}:
Calibtools: Posterior Probability Calibration Tools for Machine Learning Algorithms.
(2021).
R package version 0.0.0.9000.
\url{https://github.com/kenneth-lee-ch/calibtools}
\end{botherref}
\endbibitem

\bibitem[\protect\citeauthoryear{Schwarz and Heider}{2019}]{CalibratR-package}
\begin{barticle}
\bauthor{\bsnm{Schwarz}, \binits{J.}},
\bauthor{\bsnm{Heider}, \binits{D.}}:
\batitle{{GUESS}: Projecting machine learning scores to well-calibrated probability estimates for clinical decision making}.
\bjtitle{Bioinformatics}
\bvolume{35}(\bissue{14}),
\bfpage{2458}--\blpage{2465}
(\byear{2019})
\end{barticle}
\endbibitem

\bibitem[\protect\citeauthoryear{Kuhn et~al.}{2023}]{probably-package}
\begin{botherref}
\oauthor{\bsnm{Kuhn}, \binits{M.}},
\oauthor{\bsnm{Vaughan}, \binits{D.}},
\oauthor{\bsnm{Ruiz}, \binits{E.}}:
Probably: Tools for Post-Processing Class Probability Estimates.
(2023).
R package version 1.0.2.
\url{https://CRAN.R-project.org/package=probably}
\end{botherref}
\endbibitem

\bibitem[\protect\citeauthoryear{Zadrozny and Elkan}{2001}]{ZadroznyElkan2001}
\begin{bchapter}
\bauthor{\bsnm{Zadrozny}, \binits{B.}},
\bauthor{\bsnm{Elkan}, \binits{C.P.}}:
\bctitle{Obtaining calibrated probability estimates from decision trees and naive bayesian classifiers}.
In: \bbtitle{ICML}
(\byear{2001})
\end{bchapter}
\endbibitem

\bibitem[\protect\citeauthoryear{Naeini et~al.}{2015}]{NaeiniCooperHauskrecht2015}
\begin{botherref}
\oauthor{\bsnm{Naeini}, \binits{M.P.}},
\oauthor{\bsnm{Cooper}, \binits{G.}},
\oauthor{\bsnm{Hauskrecht}, \binits{M.}}:
Obtaining well calibrated probabilities using bayesian binning.
Proceedings of the AAAI Conference on Artificial Intelligence
\textbf{29}(1)
(2015)
\doiurl{10.1609/aaai.v29i1.9602}
\end{botherref}
\endbibitem

\bibitem[\protect\citeauthoryear{Gonzalez and Wu}{1999}]{Gonzalez1999}
\begin{barticle}
\bauthor{\bsnm{Gonzalez}, \binits{R.}},
\bauthor{\bsnm{Wu}, \binits{G.}}:
\batitle{On the shape of probability weighting function}.
\bjtitle{Cognitive Psychology}
\bvolume{38},
\bfpage{129}--\blpage{66}
(\byear{1999})
\doiurl{10.1006/cogp.1998.0710}
\end{barticle}
\endbibitem

\bibitem[\protect\citeauthoryear{Turner et~al.}{2014}]{Turner2014}
\begin{barticle}
\bauthor{\bsnm{Turner}, \binits{B.}},
\bauthor{\bsnm{Steyvers}, \binits{M.}},
\bauthor{\bsnm{Merkle}, \binits{E.}},
\bauthor{\bsnm{Budescu}, \binits{D.}},
\bauthor{\bsnm{Wallsten}, \binits{T.}}:
\batitle{Forecast aggregation via recalibration}.
\bjtitle{Machine Learning}
\bvolume{95},
\bfpage{261}--\blpage{289}
(\byear{2014})
\doiurl{10.1007/s10994-013-5401-4}
\end{barticle}
\endbibitem

\bibitem[\protect\citeauthoryear{Kass and Raftery}{1995}]{KassRaftery1995}
\begin{barticle}
\bauthor{\bsnm{Kass}, \binits{R.E.}},
\bauthor{\bsnm{Raftery}, \binits{A.E.}}:
\batitle{Bayes factors}.
\bjtitle{Journal of the American Statistical Association}
\bvolume{90}(\bissue{430}),
\bfpage{773}--\blpage{795}
(\byear{1995})
\doiurl{10.1080/01621459.1995.10476572}
\end{barticle}
\endbibitem

\bibitem[\protect\citeauthoryear{Kass and Wasserman}{1995}]{Kass1995Reference}
\begin{barticle}
\bauthor{\bsnm{Kass}, \binits{R.E.}},
\bauthor{\bsnm{Wasserman}, \binits{L.}}:
\batitle{A reference bayesian test for nested hypotheses and its relationship to the schwarz criterion}.
\bjtitle{Journal of the American Statistical Association}
\bvolume{90}(\bissue{431}),
\bfpage{928}--\blpage{934}
(\byear{1995})
\doiurl{10.1080/01621459.1995.10476592}
{\href{https://arxiv.org/abs/https://www.tandfonline.com/doi/pdf/10.1080/01621459.1995.10476592}{{https://www.tandfonline.com/doi/pdf/10.1080/01621459.1995.10476592}}}
\end{barticle}
\endbibitem

\bibitem[\protect\citeauthoryear{Nelder and Mead}{1965}]{nelder-mead}
\begin{barticle}
\bauthor{\bsnm{Nelder}, \binits{J.A.}},
\bauthor{\bsnm{Mead}, \binits{R.}}:
\batitle{{A Simplex Method for Function Minimization}}.
\bjtitle{The Computer Journal}
\bvolume{7}(\bissue{4}),
\bfpage{308}--\blpage{313}
(\byear{1965})
\doiurl{10.1093/comjnl/7.4.308}
{\href{https://arxiv.org/abs/https://academic.oup.com/comjnl/article-pdf/7/4/308/1013182/7-4-308.pdf}{{https://academic.oup.com/comjnl/article-pdf/7/4/308/1013182/7-4-308.pdf}}}
\end{barticle}
\endbibitem

\bibitem[\protect\citeauthoryear{Nocedal and Wright}{1999}]{NocedalWright1998}
\begin{bbook}
\bauthor{\bsnm{Nocedal}, \binits{J.}},
\bauthor{\bsnm{Wright}, \binits{S.J.}}:
\bbtitle{Numerical Optimization}.
\bpublisher{Springer}, \blocation{???}
(\byear{1999})
\end{bbook}
\endbibitem

\bibitem[\protect\citeauthoryear{Johnson}{2007}]{nlopr-package}
\begin{botherref}
\oauthor{\bsnm{Johnson}, \binits{S.G.}}:
The {NLopt} nonlinear-optimization package.
\url{https://github.com/stevengj/nlopt}
(2007)
\end{botherref}
\endbibitem

\bibitem[\protect\citeauthoryear{Kraft}{1988}]{slsqp1}
\begin{botherref}
\oauthor{\bsnm{Kraft}, \binits{D.}}:
A software package for sequential quadratic programming
\textbf{88},
1--33
(1988)
\end{botherref}
\endbibitem

\bibitem[\protect\citeauthoryear{Kraft}{1994}]{slsqp2}
\begin{barticle}
\bauthor{\bsnm{Kraft}, \binits{D.}}:
\batitle{Algorithm 733: Tomp–fortran modules for optimal control calculations}.
\bjtitle{ACM Trans. Math. Softw.}
\bvolume{20}(\bissue{3}),
\bfpage{262}--\blpage{281}
(\byear{1994})
\doiurl{10.1145/192115.192124}
\end{barticle}
\endbibitem

\bibitem[\protect\citeauthoryear{Conn et~al.}{1991}]{auglag1}
\begin{barticle}
\bauthor{\bsnm{Conn}, \binits{A.R.}},
\bauthor{\bsnm{Gould}, \binits{N.I.M.}},
\bauthor{\bsnm{Toint}, \binits{P.L.}}:
\batitle{A globally convergent augmented lagrangian algorithm for optimization with general constraints and simple bounds}.
\bjtitle{SIAM J. Numer. Anal.}
\bvolume{28}(\bissue{2}),
\bfpage{545}--\blpage{572}
(\byear{1991})
\doiurl{10.1137/0728030}
\end{barticle}
\endbibitem

\bibitem[\protect\citeauthoryear{Birgin and Mart{\'i}nez}{2008}]{auglag2}
\begin{barticle}
\bauthor{\bsnm{Birgin}, \binits{E.G.}},
\bauthor{\bsnm{Mart{\'i}nez}, \binits{J.M.}}:
\batitle{Improving ultimate convergence of an augmented lagrangian method}.
\bjtitle{Optimization Methods and Software}
\bvolume{23},
\bfpage{177}--\blpage{195}
(\byear{2008})
\end{barticle}
\endbibitem

\bibitem[\protect\citeauthoryear{Svanberg}{2002}]{MMA}
\begin{barticle}
\bauthor{\bsnm{Svanberg}, \binits{K.}}:
\batitle{A class of globally convergent optimization methods based on conservative convex separable approximations}.
\bjtitle{SIAM Journal on Optimization}
\bvolume{12},
\bfpage{555}--\blpage{573}
(\byear{2002})
\doiurl{10.1137/S1052623499362822}
\end{barticle}
\endbibitem

\bibitem[\protect\citeauthoryear{{Douglas Nychka} et~al.}{2021}]{fields-package}
\begin{botherref}
\oauthor{\bsnm{{Douglas Nychka}}},
\oauthor{\bsnm{{Reinhard Furrer}}},
\oauthor{\bsnm{{John Paige}}},
\oauthor{\bsnm{{Stephan Sain}}}:
fields: Tools for spatial data,
Boulder, CO, USA.
R package version 15.2
(2021).
\url{https://github.com/dnychka/fieldsRPackage}
\end{botherref}
\endbibitem

\bibitem[\protect\citeauthoryear{Wickham}{2016}]{ggplot2-package}
\begin{bbook}
\bauthor{\bsnm{Wickham}, \binits{H.}}:
\bbtitle{Ggplot2: Elegant Graphics for Data Analysis}.
\bpublisher{Springer},
\blocation{New York}
(\byear{2016}).
\burl{https://ggplot2.tidyverse.org}
\end{bbook}
\endbibitem

\bibitem[\protect\citeauthoryear{FiveThirtyEight}{2021}]{fivethirtyeight}
\begin{botherref}
\oauthor{\bsnm{FiveThirtyEight}}:
2020-21 NHL Predictions.
Accessed: 2022-01-05.
\url{https://data.fivethirtyeight.com/}
Accessed 2022-11-08
\end{botherref}
\endbibitem

\bibitem[\protect\citeauthoryear{Keefe et~al.}{2017}]{Keefe2017Monitoring}
\begin{barticle}
\bauthor{\bsnm{Keefe}, \binits{M.J.}},
\bauthor{\bsnm{Franck}, \binits{C.T.}},
\bauthor{\bsnm{Woodall}, \binits{W.H.}}:
\batitle{Monitoring foreclosure rates with a spatially risk-adjusted bernoulli cusum chart for concurrent observations}.
\bjtitle{Journal of Applied Statistics}
\bvolume{44}(\bissue{2}),
\bfpage{325}--\blpage{341}
(\byear{2017})
\doiurl{10.1080/02664763.2016.1169257}
\end{barticle}
\endbibitem

\bibitem[\protect\citeauthoryear{Powell}{1994}]{cobyla}
\begin{bbook}
\bauthor{\bsnm{Powell}, \binits{M.J.D.}}:
In: \beditor{\bsnm{Gomez}, \binits{S.}},
\beditor{\bsnm{Hennart}, \binits{J.-P.}} (eds.)
\bbtitle{A Direct Search Optimization Method That Models the Objective and Constraint Functions by Linear Interpolation},
pp. \bfpage{51}--\blpage{67}.
\bpublisher{Springer},
\blocation{Dordrecht}
(\byear{1994}).
\doiurl{10.1007/978-94-015-8330-5_4} .
\burl{https://doi.org/10.1007/978-94-015-8330-5_4}
\end{bbook}
\endbibitem

\end{thebibliography}

\end{document}